\documentclass[sn-nature]{sn-jnl}

\usepackage{graphicx}%
\usepackage{multirow}%
\usepackage{amsmath,amssymb,amsfonts}%
\usepackage{amsthm}%
\usepackage{mathrsfs}%
\usepackage[title]{appendix}%
\usepackage{xcolor}%
\usepackage{textcomp}%
\usepackage{manyfoot}%
\usepackage{booktabs}%
\usepackage{algorithm}%
\usepackage{algorithmicx}%
\usepackage{algpseudocode}%
\usepackage{listings}%
\usepackage{subcaption}%
\usepackage[export]{adjustbox}%

\theoremstyle{thmstyleone}%

\theoremstyle{thmstyletwo}%

\theoremstyle{thmstylethree}%

\begin{document}

\title[Article Title]{Substrate-aware computational design of two-dimensional materials}

\author*[1]{\fnm{Arslan} \sur{Mazitov}}\email{arslan.mazitov@phystech.edu}

\author[1,2]{\fnm{Ivan} \sur{Kruglov}}

\author[1]{\fnm{Alexey V.} \sur{Yanilkin}}

\author[1,2,3]{\fnm{Aleksey V.} \sur{Arsenin}}

\author[2,3]{\fnm{Valentyn S.} \sur{Volkov}}

\author[4]{\fnm{Dmitry G.} \sur{Kvashnin}}

\author[5]{\fnm{Artem R.} \sur{Oganov}}

\author[6,7,8]{\fnm{Kostya S.} \sur{Novoselov}}

\affil[1]{\orgname{Moscow Center for Advanced Studies}, \orgaddress{\street{Kulakova str. 20}, \city{Moscow}, \postcode{123592}, \country{Russian Federation}}}

\affil[2]{\orgdiv{Emerging Technologies Research Center}, \orgname{XPANCEO}, \orgaddress{\street{Internet City, Emmay Tower}, \city{Dubai},  \country{United Arab Emirates}}}

\affil[3]{\orgdiv{Laboratory of Advanced Functional Materials}, \orgname{Yerevan State University}, \city{Yerevan}, \postcode{0025}  \country{Armenia}}

\affil[4]{\orgname{Emanuel Institute of Biochemical Physics}, \orgaddress{ \street{Kosigina st. 4}  \city{Moscow}, \postcode{119334},  \country{Russian Federation}}}

\affil[5]{\orgdiv{Materials Discovery Laboratory}, \orgname{Skolkovo Institute of Science and Technology (Skoltech)}, \orgaddress{\street{Bolshoy Boulevard 30, bld. 1}, \city{Moscow}, \postcode{121205},  \country{Russian Federation}}}

\affil[6]{\orgdiv{National Graphene Institute (NGI)}, \orgname{University of Manchester}, \orgaddress{\city{Manchester}, \postcode{M13 9PL},  \country{UK}}}

\affil[7]{\orgdiv{Department of Materials Science and Engineering}, \orgname{National University of Singapore}, \orgaddress{\city{Singapore}, \postcode{03-09 EA},  \country{Singapore}}}

\affil[8]{\orgdiv{Institute for Functional Intelligent Materials}, \orgname{National University of Singapore}, \orgaddress{\city{Singapore}, \postcode{117544},  \country{Singapore}}}

\abstract{Two-dimensional (2D) materials have attracted considerable attention due to their remarkable electronic, mechanical and optical properties, making them prime candidates for next-generation electronic and optoelectronic applications. Despite their widespread use in combination with substrates in practical applications, including the fabrication process and final device assembly, computational studies often neglect the effects of substrate interactions for simplicity. This study presents a novel method for predicting the atomic structure of 2D materials on arbitrary substrates by combining an evolutionary algorithm, a lattice-matching technique, an automated machine learning interatomic potentials training protocol, and the \textit{ab initio} thermodynamics approach for predicting the possible conditions of experimental synthesis of the predicted 2D structures. Our method allows the automatic exploration of atomic configurations and chemical compositions, assesses their stability in the presence of a substrate, and provides phase diagrams of the predicted structures in the parameter space of a desired experimental setup. Using the molybdenum-sulfur (Mo-S) system on a $c$-cut sapphire (Al$_2$O$_3$) substrate as a case study, we demonstrate the reliability of our technique. The evolutionary search revealed several new stable and metastable structures, including previously known 1H-MoS$_2$ and newly found $Pmma$ Mo$_3$S$_2$, $P\bar{1}$ Mo$_2$S, $P2_1m$ Mo$_5$S$_3$, and $P4mm$ Mo$_4$S, where the Mo$_4$S structure is specifically stabilized by interaction with the substrate. Electronic band structure calculations of Mo$_3$S$_2$, Mo$_2$S, Mo$_5$S$_3$, and Mo$_4$S showed their metallic behavior, while phonon properties calculations indicated their dynamic stability and substrate-induced modulation of the phonon density of states. Finally, we use the ab initio thermodynamics approach to predict the synthesis conditions of the discovered structures in the parameter space of the commonly used chemical vapor deposition technique. These results provide insights into computational substrate engineering, allowing one to study the substrate effect on the thermodynamic and dynamical stability of 2D materials and to modulate their electronic and phonon properties for their future applications, as well as to provide guide maps for their experimental synthesis.}


\maketitle

\section{Introduction}\label{sec:intro}

The discovery of graphene, a single layer of carbon atoms arranged in a honeycomb lattice, in 2004 marked a milestone in the field of 2D materials and sparked a wave of research into this novel class of materials \cite{Novoselov2004ElectricFilms}. Since then, the family of 2D materials has expanded to include transition metal dichalcogenides (TMDs), hexagonal boron nitride (h-BN), phosphorene, and other layered materials, each with their own unique properties and potential applications \cite{Wang2012ElectronicsDichalcogenides, Castellanos-Gomez2014IsolationPhosphorus, Naclerio2023AApplications}. Moreover, this family is only getting larger \cite{Anasori2024AcceleratingDiscovery}, and many new 2D materials, such as Ru$_2$Si$_x$O$_y$ , Cr$_2$B$_2$F$_2$ and Ni$_2$Si$_2$O$_2$ have recently been discovered by large-scale computational studies, while the stability of Ru$_2$Si$_x$O$_y$ has been confirmed by further experimental synthesis \cite{Bjork2024Two-dimensionalSolids}. At the time of writing, the Computational 2D Materials Database \cite{Haastrup2018TheCrystals, Gjerding2021RecentC2DB} already contains more than 16,000 entries. 

The unique properties of 2D materials, such as high carrier mobility, exceptional mechanical flexibility, and tunable bandgaps, have spurred interest in exploring their use in a wide range of electronic and optoelectronic devices \cite{Novoselov2005Two-dimensionalGraphene, Wang2012ElectronicsDichalcogenides, Gupta2015RecentGraphene, Shao2017AClustering, Manzeli20172DDichalcogenides, Iannaccone2018QuantumHeterostructures}.  For example, TMDs have emerged as promising candidates for field-effect transistors, photodetectors, and light-emitting diodes due to their large bandgaps and strong light-matter interaction \cite{Wang2012ElectronicsDichalcogenides}. Similarly, the atomic thinness and excellent mechanical properties of graphene make it an ideal candidate for flexible electronics, transparent conductive electrodes, and sensors \cite{Novoselov2012AGraphene}. Moreover, a wide range of ways to tune properties of 2D layers using lateral and vertical heterostructures fabrication \cite{Geim2013VanHeterostructures,Novoselov20162DHeterostructures,Liu2016VanDevices}, chemical functionalization \cite{Gupta2015RecentGraphene,Liu2015FunctionalizedProcessing}, strain \cite{Ahn2017Strain-engineeredMaterials, Yan2018TowardMethods, Wu2022StrainEngineering}, defect \cite{Hus2017Spatially-resolvedMaterials,Rhodes2019DisorderMaterials} and substrate engineering \cite{Shao2017AClustering}, makes them ideal candidates for developing a new class of electronic devices. According to International Technology Roadmap for Semiconductors \cite{InternationalEdition}, the use of 2D materials and their heterostructures in the fabrication of a new generation of transistors can improve the technological process from $\sim$ 5 nm to 1.5 - 2 nm by 2030. Moreover, many promising applications in the fields of photonics  \cite{Mak2016PhotonicsDichalcogenides}, photovoltaics \cite{Liu2015FunctionalizedProcessing, Das2019ThePhotovoltaics}, valleytronics \cite{Schaibley2016ValleytronicsMaterials}, energetics \cite{Shi2017RecentBatteries}, and catalysis \cite{Luo2016RecentPhotocatalysis} have already been realized in practice. 

2D materials provide a very versatile platform to study a large range of physical phenomena. However, even more control can be achieved through the manipulation of the interaction with the substrate. Indeed, the fabrication of 2D materials typically relies on physical (PVD) and chemical (CVD) vapor deposition techniques, molecular-beam (MBE) and atomic-layer (ALE) epitaxy, or direct mechanical exfoliation method \cite{Iannaccone2018QuantumHeterostructures,Mannix2017SynthesisMaterials}, where the material is either directly grown or finally placed on top of a substrate. Substrate engineering not only offers the higher quality of the devices, as in the case of graphene on hBN \cite{Dean2012GrapheneHeterostructures, Kim2012SynthesisDevices}, but also allows to modify the stability, electronic structure, and mechanical properties of the 2D material-substrate system. For example, the choice of substrate can induce strain in the 2D material, thereby modulating its electronic properties and band structure \cite{Liu2016VanDevices, Rhodes2019DisorderMaterials}. The interaction with the substrate can also lead up to significant reduction in graphene's thermal conductivity on $\alpha$-SiO$_2$ compared to freestanding graphene due to enhanced phonon-phonon scattering rate  \cite{Shao2017AClustering}.  In addition, the substrate can influence charge transfer processes \cite{Wang2019SubstrateMaterials}, affect device performance, induce semi-metal - metal transition \cite{Fan2018SiliceneProperties}, or even lead to the emergence of novel phenomena such as moiré patterns in heterostructures \cite{Yankowitz2019VanNitride}.  

In computational 2D materials design, substrate effects however are often overlooked for the sake of simplicity. Most of the studies based on high-throughput screening of 2D materials with machine learning and/or \textit{ab initio} techniques typically consider isolated 2D materials without taking into account the presence of substrates or other surrounding layers \cite{Rashid2024ReviewDesign}. This includes data-driven high-throughput \textit{ab initio} screening of 2D materials for properties optimization  \cite{Sarikurt2020High-throughputThermoelectrics, Zhang2019High-throughputMaterials, Mounet2018Two-dimensionalCompounds, Gjerding2021RecentC2DB, Zhuang2014ComputationalMaterials},  evolutionary and global optimization methods for 2D materials discovery \cite{Zhou2014SemimetallicFermions, Dong2019NewMaterials, Miao2020ComputationalDerivatives, Popov2021NovelAlgorithm, Zhou2016Two-dimensionalBoron, Gu2017PredictionApproach, Borlido2017StructuralStrain, Wang2012AnAlgorithm,Penev2021TheoreticalProperties, Zhuang2014ComputationalMaterials, Revard2016Grand-canonicalMaterials}, and machine learning based prediction of new 2D materials \cite{Dong2023DiscoveryDesign, Zeni2023MatterGen:Design, Song2021ComputationalModels, Agarwal2021Data-DrivenSplitting}. In the case of crystal structure prediction (CSP), predicting the structure of two-dimensional materials in a wide range of stochiometires and in presence of the substrate at the \textit{ab initio} accuracy was desired, yet previously inaccessible for a few reasons. First, the periodic representation of the 2D layer and the substrate requires their unit cells to be as close as possible to each other to avoid the unphysical mistfit stress \cite{Singh2014AbMaterials}. Since this is not the case for primitive cells in general, a suitable supercell construction is usually required to minimize the lattice mismatch. This inevitably leads to the structures with a large number of atoms, which does not allow one to use conventional density functional theory (DFT) calculations.

This paper introduces a method for substrate-aware computational design of 2D materials. The method is based on a combination of the evolutionary algorithm USPEX \cite{Oganov2006CrystalApplications, Oganov2011HowWhy, Lyakhov2013NewUSPEX}, a lattice-matching technique, a machine learning interatomic potentials (MLIP) relaxation protocol, and the \textit{ab initio} thermodynamics approach to study the possible conditions for experimental synthesis of predicted 2D structures. It allows for the automatic exploration of the full range of atomic configurations and chemical compositions for given elements, as well as the estimation of their relative stability in the presence of the substrate, and linking the calculated stability patterns to a realistic set of parameters controlled by a specific experimental setup. As the training of the MLIP for CSP purposes represents a crucial and non-trivial part of the computational pipeline, we also introduce an automatic workflow for creating a robust interatomic potential suitable for CSP. We demonstrate the reliability of our technique by predicting the stable two-dimensional crystals in the molybdenum-sulfur system on a $c$-cut sapphire substrate. First, we create the interatomic potential for the Mo-S @ Al$_2$O$_3$ system, which is suitable for both predicting the structure of freestanding two-dimensional molybdenum-sulfur layers and those joined with the substrate. Subsequently, an evolutionary search for stable two-dimensional crystals is performed using the trained machine learning potential for local relaxation and stability evaluation of the Mo-S structures. Finally, for all the stable structures, we predict the conditions of their synthesis in the parameters space of the commonly used CVD technique.

\section{Automatic self-consistent training of MLIPs}\label{sec:asct}

To train the interatomic potential used in our work, we developed an automatic self-consistent training (ASCT) algorithm based on iterative sampling of new structures from molecular dynamics (MD) trajectories combined with subsequent \textit{ab initio} calculations of their energies, interatomic forces, and stresses. The use of a MLIP offers several advantages, such as a reasonable accuracy of the energy and force evaluations, together with a significantly lower computational cost, which allows one to consider large structures with low values of lattice mismatch. However, the train database sampling for such a potential typically requires a lot of effort due to the large space of different configurations that need to be covered. The potential should be able to handle the random 2D crystal structures generated within the evolutionary algorithm (EA) and perform their structural optimization together with MD annealing at constant pressure and temperature. This includes both the free-standing 2D structures, those connected to the substrate, and the free-standing surface slab lattice dynamics due to the relaxation technique used in our work (see section \ref{sec:ea_details} for details).

\begin{figure}
    \centering
    \includegraphics[width=\columnwidth]{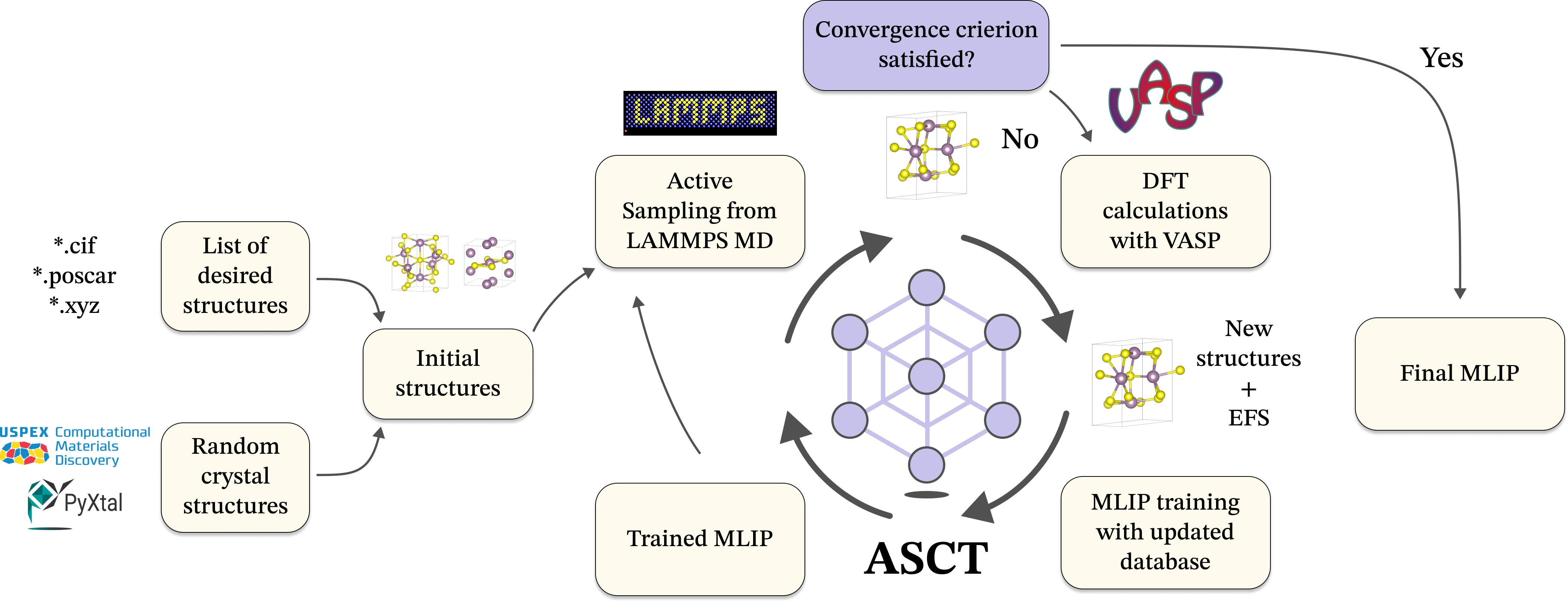}
    \caption{Automatic self-consistent training workflow (ASCT). At each iteration of the ASCT, a set of candidate structures is selected from the MD trajectory initialized with either randomly generated crystals or a given list of desired atomic configurations. New structures are sampled using the ML potential from the previous iteration based on the extrapolation criterion. These structures are then used to update the training set of the potential after calculating their energies, forces and stresses in the DFT, and the training cycle is repeated. Finally, the training stops if no new structures are sampled from the MD runs for a certain number of cycles.}
    \label{fig:asct}
\end{figure}

The schematic representation of ASCT is given in a Figure \ref{fig:asct}. At each iteration of the ASCT, a number of seed atomic configurations are either generated by the random structure generator of the USPEX code or selected from a list of structures provided by the user. These configurations are then used to initialize the parallel MD sampling of new configurations within the LAMMPS package \cite{Plimpton1995FastDynamics, Thompson2022LAMMPSScales} using the ML potential from the previous training iteration. During the MD run, each structure is compared to the existing training set of the ML potential, and is saved if it gets beyond the corresponding region of the configuration space. While the potential is not fully trained, the sampling procedure is usually interrupted after reaching a certain extrapolation criteria of the ML potential to avoid sampling unphysical configurations. All configurations from parallel MD runs are then aggregated, existing duplicates are removed from the list, and energies, interatomic forces and stresses of the remaining configurations are calculated with the VASP package. Finally, the new data set is merged with the training set from the previous iteration and the ML potential is retrained. Once the configuration space of a given system is fully explored during the ASCT and the new configurations are not sampled during the MD runs for a reasonable number of iterations (usually about 10-15), the convergence of the ASCT is considered to be achieved. The first iteration of the ASCT is typically initialized with a ML potential pre-trained on a few representative configurations, which can be either prepared manually or sampled from the short \textit{ab initio} MD trajectory. Our ASCT workflow is currently interfaced only with the Moment Tensor Potentials (MTP) framework \cite{Shapeev2016MomentPotentials}, which was chosen as the primary ML potential implementation in this work because of its relatively high accuracy \cite{Zuo2020PerformancePotentials, Lysogorskiy2021PerformantSilicon}, convenient code implementation \cite{Novikov2020TheLearning}, and active learning capabilities \cite{Podryabinkin2023MLIP-3:Potentials}. Moreover, it has been successfully used in many recent computational studies for predicting the crystal structure of bulk materials \cite{Podryabinkin2019AcceleratingLearning}, phase behavior of alloys \cite{Gubaev2019AcceleratingPotentials, Kostiuchenko2019ImpactPotentials, Kostiuchenko2020Short-rangeAlloys, Nyshadham2019Machine-learnedPrediction}, and characterizing the phonon, kinetic and mechanical properties of 2D materials \cite{Mortazavi2020ExploringPotentials, Mortazavi2020Machine-learningHeterostructures, Mortazavi2021First-PrinciplesPotentials}. 

Using the ASCT framework, we have thus trained the ML potential for the 2D Mo-S/Al$_2$O$_3$ system, which is suitable for crystal structure prediction purposes. Since the computational cost is much lower compared to DFT calculations, the potential is able to calculate energies, forces and stresses for large structures, thus allowing to work with large cells with low lattice mismatch. We provide the technical details of the training in Sec. \ref{sec:mlip}, while only the final results are demonstrated in this part of the manuscript. Table \ref{tab:mlip} shows the accuracy of the final model in predicting energies, interatomic forces and stresses of the structures in the training set. When interpreting these results, it should be kept in mind that (1) the training set consists mostly of random, and thus highly diverse and non-equilibrium structures; (2) even though the given accuracy is calculated on the training data, it is also relevant for the real scenario test applications, since our training technique effectively covers all possible simulation scenarios in the evolutionary search; (3) the percentage values of the root mean square error are derived as the ratio between the RMS deviation of a quantity from its mean and the RMS value of the quantity, essentially giving an insight into how large the errors are compared to the mean value of the quantity. In Supplementary Figure 1, we also provide training and validation energy pair plots to evaluate the performance of the potential in the extrapolation regime (i.e., on the structures that were not part of the training set).

\begin{table}[h!]
\centering
 \begin{tabular}{|c | c | c | c|} 
 \hline
  & Energies, meV/at. & Forces, eV/\AA \ &  Stresses, kbar\\
 \hline
 MAE & 16.7 & 0.89 & 2.38\\
 RMSE& 26.3& 1.45&4.25\\
 \hline
 RMSE, \% & \hfill & 12.5 & 2.7 \\
 \hline
Total number of configurations &  \multicolumn{3}{c|}{3240} \\
\hline
Total number of atoms in all configurations&  \multicolumn{3}{c|}{283059} \\
\hline
 \end{tabular}
 \caption{Performance of the trained MTP potential for the Mo-S/Al$_2$O$_3$ system.}
 \label{tab:mlip}
\end{table}

Even though the errors presented in Table \ref{tab:mlip} are not negligible, the trained MLIP can be used for high-throughput screening of potentially stable candidates in a given system, with subsequent refinement of the results within accurate DFT calculations. Nevertheless, as one can see in the EA results below (section \ref{sec:ea}), the convex hulls given by MTP and recalculated with DFT turn out to be quite close. This may indicate that the training error accumulates on highly non-equilibrium structures, while the predictions on relaxed structures are more accurate. 

\section{Substrate-aware crystal structure prediction of 2D materials}\label{sec:ea}

While predicting the structure of free-standing 2D crystals is a common practice \cite{Zhou2014SemimetallicFermions, Dong2019NewMaterials, Miao2020ComputationalDerivatives, Popov2021NovelAlgorithm, Zhou2016Two-dimensionalBoron, Gu2017PredictionApproach, Borlido2017StructuralStrain, Wang2012AnAlgorithm,Penev2021TheoreticalProperties}, incorporating substrate effects into this type of simulation has been challenging for several reasons. First, the primitive cells of a 2D crystal and a substrate are usually not exactly the same, which leads to the non-physical strain caused by lattice mismatch. One of the possible solutions here is to find a suitable supercell representation that matches both the 2D crystal and the substrate, thus reducing the mismatch effect on the structure and energy of the considered material. Second, the resulting supercell size of 10$^3$-10$^4$ atoms does not  allow one to use conventional DFT codes to perform hundreds of structural relaxations and energy evaluations of the crystals that are typically required during the EA run. Machine learning based models, which usually solve the puzzle by combining near \textit{ab initio} accuracy and low computational cost, may indeed remain the only available option, but require a thorough preparation of the training data for each particular system. In this section, and in section \ref{sec:asct}, we address both problems and propose a robust and convenient way both to perform the CSP and to train the required ML model.

The evolutionary search for the new stable structures in a 2D Mo-S/Al$_2$O$_3$ system was performed with a modified version of the USPEX code.  USPEX has been widely used  \cite{Oganov2019StructureDiscovery} for predicting the crystal structure of bulk materials, two-dimensional materials in a vacuum \cite{Zhou2016Two-dimensionalBoron}, reconstructions of surfaces \cite{Zhu2013EvolutionaryStoichiometry} and nanoclusters \cite{Lepeshkin2018MethodCompositions}, as well as for discovering stable materials with optimal properties \cite{Nunez-Valdez2018EfficientMaterials, Kvashnin2017ComputationalMaterials}

\begin{figure}
    \centering
    \includegraphics[width=\columnwidth]{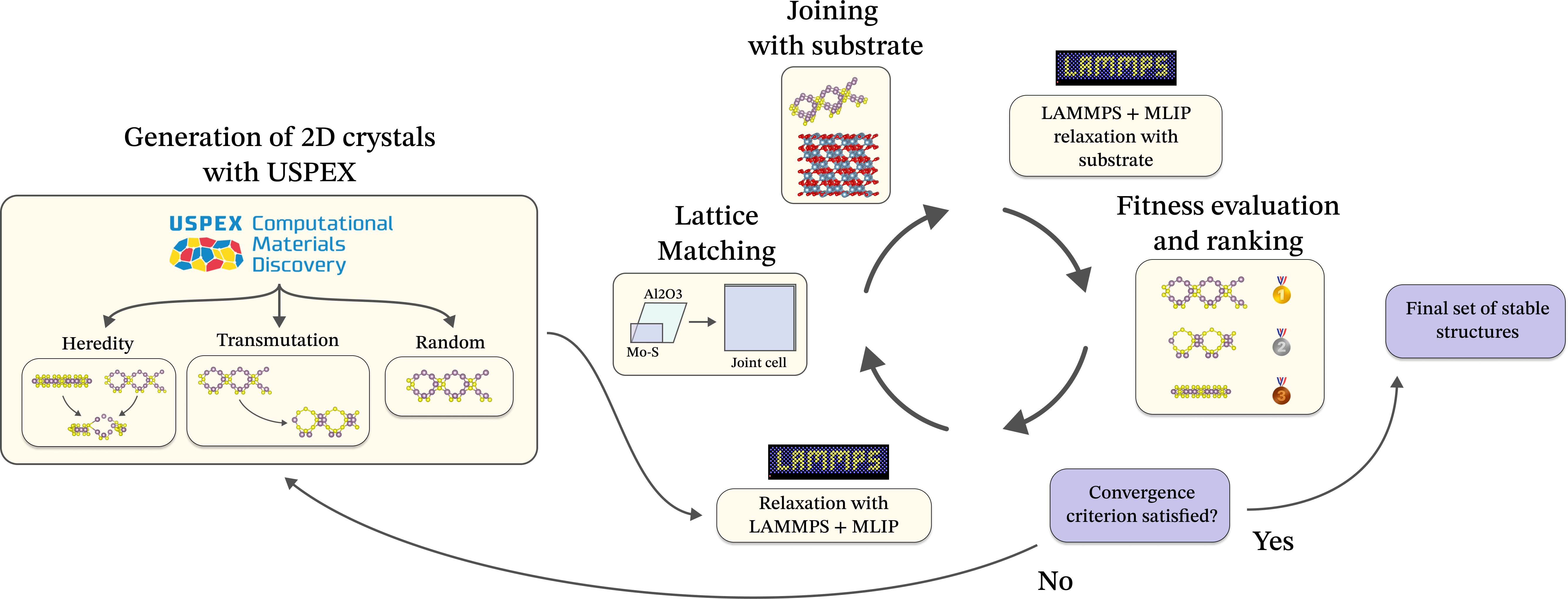}
    \caption{Evolutionary algorithm workflow. At each iteration, a set of 2D structures are generated using the evolutionary operators in a given range of chemical compounds. After a preliminary relaxation of the atomic positions and lattice parameters, these structures are attached to the substrate using a lattice-matching algorithm that ensures no lattice mismatch. Finally, the structures are ranked by their fitness value and the cycle is repeated. The calculation stops when the set of best structures remains unchanged for a given number of iterations. Detailed explanation of the algorithm and evolutionary operators used is given in Methods, Sec.  \ref{sec:ea_details}.}
    \label{fig:ea_workflow}
\end{figure}

The workflow of the algorithm developed is summarized in Figure \ref{fig:ea_workflow}. In each generation, a set of 2D crystal structures with a constrained thickness is first produced with either USPEX evolutionary operators, or using a random structure generator \cite{Bushlanov2019Topology-basedGenerator, Fredericks2021PyXtal:Analysis}. These structures then undergo a preliminary relaxation and annealing step without a substrate with the trained ML potential. The reason why we relax the freestanding crystals before joining them with the substrate and not vice versa is that we assume the equilibrium state of a crystal to be mostly determined by its internal composition, while the effect of the substrate is considered as a next-order correction. As the equilibrium lattice parameters and atomic positions are determined, we join the 2D crystal with the substrate using a Lattice Matching algorithm (see Sec. \ref{sec:lattice_matching} for details) and perform a conjugate gradient relaxation of the atomic positions, followed by evaluation of a total energy of the system. To subtract the substrate contribution, we detach the substrate and calculate its energy while keeping atoms fixed. Finally, each structure in the set is ranked according to its value of fitness, which is calculated as a height above the convex hull in a Mo-S composition space. We note, that the convex hull construction is valid here precisely for the reason of fixed thickness (see more details in Ref. \cite{Revard2016Grand-canonicalMaterials}). The best representatives of the current set of structures are then used in the next generation of EA to produce the new set of structures using the heredity and transmutation evolutionary operators (see Sec. \ref{sec:ea_details} for details). Thus the algorithm works iteratively until the list of best structures remains unchanged for a reasonable amount of generations. 

\section{Stable 2D Mo-S crystals on Al$_2$O$_3$ substrate}\label{sec:ea_results}

2D TMDs are typically manufactured using CVD techniques, in which tiny flakes of the material are condensed from the gas phase of the precursors onto the surface of the substrate, where they undergo a chemical reaction. Sapphire substrates are often chosen for their durability and high chemical stability \cite{Dumcenco2015Large-areaMoS2}. Thus, we follow the outlined experimental setup and predict the stable Mo-S crystals on the same $c$-cut sapphire substrate. Our main motivation here is to computationally explore the possible space of CVD growth scenarios under varying growth parameters (e.g., partial pressures and temperatures of precursors), which may lead to changes in the chemical potentials of the elements and thus stabilize different 2D Mo-S phases. 

\begin{figure}
    \centering
    \includegraphics[width=\columnwidth]{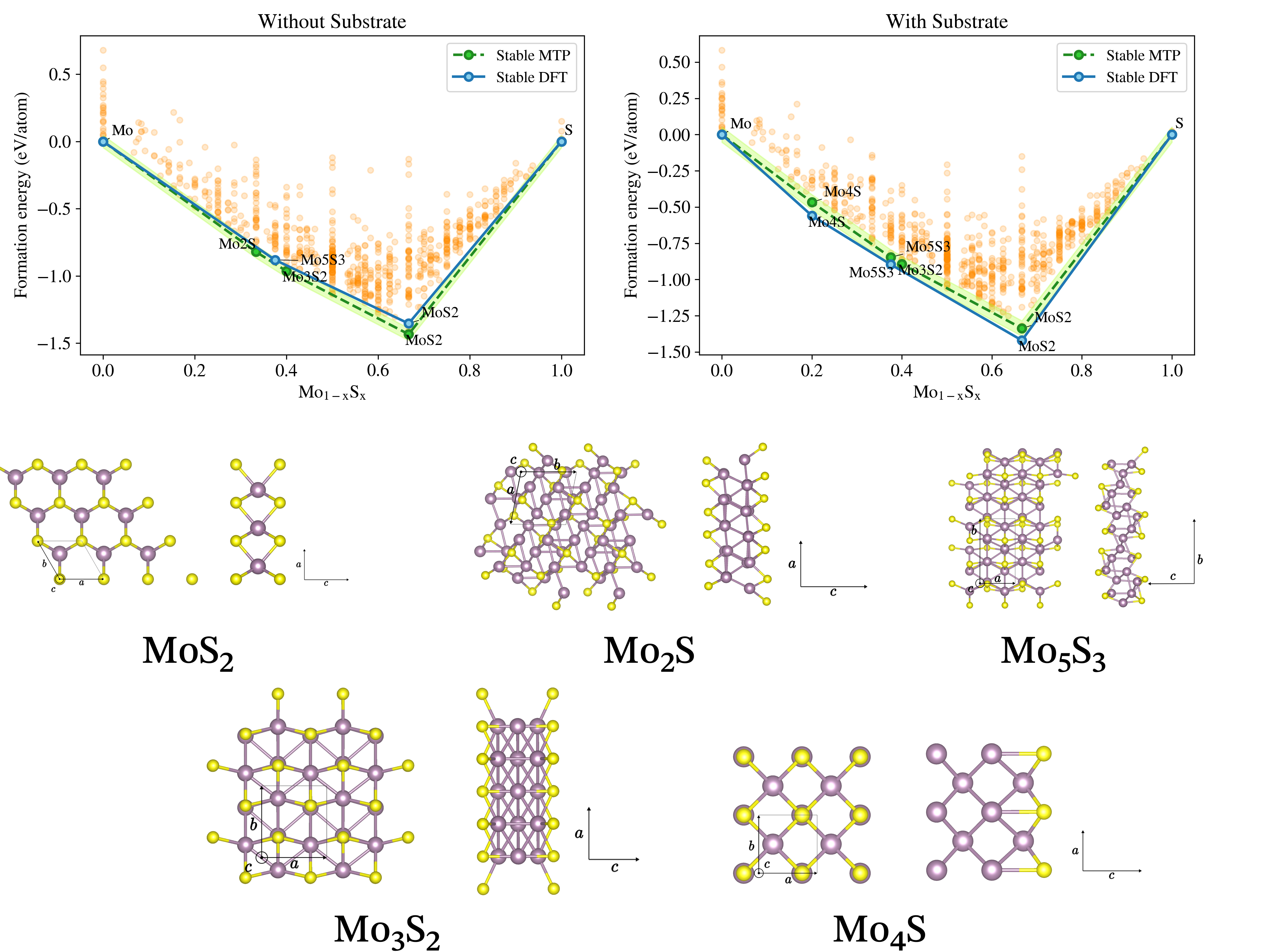}
    \caption{Results of the evolutionary search in the Mo-S/Al$_2$O$_3$ system. The upper left panel shows the convex hull together with the set of stable (green and blue circles) and metastable (orange circles) free-standing 2D Mo-S structures. The upper right panel shows similar results, but for the structures attached to the substrate. The green and blue convex hulls show the results obtained in the MTP and DFT calculations. The shaded green area represents the error of the MTP potential in predicting the energies of formation. The visualization of the convex hull structures is shown in the lower panel, where molybdenum atoms are drawn as larger violet spheres and sulfur as smaller yellow spheres. }
    \label{fig:ea_results}
\end{figure}

The results of the evolutionary search for stable Mo-S crystals on the Al$_2$O$_3$ substrate (see the details of the calculation in section \ref{sec:ea_details}) are shown in Fig. \ref{fig:ea_results}. To understand the relative thermodynamic stability of phases in binary systems, a convex hull approach in the formation energy-composition space is often used \cite{Oganov2006CrystalApplications, Oganov2011HowWhy}. The convex hull, by definition, connects phases that have lower energy than any other phase or any linear combination of phases at the same overall composition. Therefore, by plotting different structures found in the evolutionary algorithm search as points in the energy-composition space, one can determine the stable structures by identifying the smallest set of points that form a convex hull.

The top left panel in Fig. \ref{fig:ea_results} shows the convex hull for the freestanding 2D crystals, while the left panel shows the convex hull for the crystals attached to the substrate. The bottom panel contains top and side views of the 2D representation of each Mo-S structure marked on the convex hull. The green shading around the convex hull segments represents the error of the MTP potential in predicting the formation energy, which is equal to $\sqrt{3\sigma_{\text{MTP}}^2}$ and $\sqrt{4 \sigma_{\text{MTP}}^2}$ for free-standing and substrate-bound structures, respectively (where $\sigma_{\text{MTP}}$ is a root-mean-squared error of the MTP potential per-atom energies predictions from Table \ref{tab:mlip}). We chose the error window according to the errors summation rule and the energy of formation formulas used (see Sec. \ref{sec:ea_details}).  For each structure falling within this error window, we re-evaluated the energies of formation in DFT (taking into account positions and cell relaxation for the free-standing 2D layers, and using only single-point DFT calculations for those connected to the substrate). 

First of all, we note that the 1H-MoS$_2$ structure with $P\bar{6}m2$ space group and a lattice parameter $a^{\mathrm{MTP}}_{\mathrm{MoS_2}} = 3.15$ \AA \ was successfully found during the search. The lattice constant of a found structure is obtained within the MTP relaxation and very close to the experimental one, which varies from 3.12 \AA \cite{Cao2012Valley-selectiveDisulphide} to 3.22 \AA \cite{Ganatra2014Few-layerSemiconductor, Subbaiah2016AtomicallyMaterial}, indicating the high quality of the ML potential used. In addition, we found three new Mo-S structures, namely $Pmma$ Mo$_3$S$_2$, low-symmetry $P\bar{1}$ Mo$_2$S , $P2_1m$ Mo$_5$S$_3$, and $P4mm$ Mo$_4$S, which are either close to the convex hull or located on the convex hull lines and therefore are stable. Lattice parameters of the new structures are listed in the Supplementary Table 1. 

Since the main focus of the study was to determine the overall effect of the substrate on the properties of 2D materials, we next focused on analyzing the way the substrate changes the stability of the structures. Indeed, the set of stable structures on a convex hull is changed under the effect of the substrate. In both MTP and DFT cases, a new Mo$_4$S structure appears on the convex hull after joining with the substrate (see the upper right panel in Fig. 3), while Mo$_2$S looses it's stability in the MTP case. The presence of structures such as Mo$_4$S on a convex hull shows that chemical bonding with the substrate saturates a certain fraction of the bonds in the 2D layer, stabilizing structures that would not be stable in a free-standing 2D case (see Supplementary Fig. 7).  Moreover, the relative energies of formation of the structures change (see Supplementary Table 2), which allows one to use the desired parameters of experimental fabrication of the discussed materials by choosing a suitable substrate. For example, the change in the slope of the convex hull sections is directly related to the change in the chemical potentials of molybdenum and sulfur, and thus to the partial concentrations of CVD precursors under which the specific phase of the material can be stabilized. A detailed study of synthesis conditions under a given range of chemical potentials is given in Sec. \ref{sec:synthesis}.

The stability results obtained within the DFT and MTP calculations are still slightly different due to the internal inaccuracy of the potential. However, it can be observed that the sets of stable structures overlap significantly. Therefore, the proposed pipeline for stability studies requires to first perform a pure MTP-based evolutionary search combined with a successive re-evaluation of the energies of formation of a subset of structures falling within the window of MLIP errors. This approach allows one to reduce the overall cost of the study by several orders of magnitude and to achieve the scaling of the system size that is generally not available in a pure DFT approach.

The last limitation we want to discuss here is the use of potential energies to analyze the relative stability of the structures. In fact, the realistic experimental setups require Gibbs formation energy calculations and the consideration of entropy effects. In this study, for simplicity, we focused on potential energy contributions and analyzed the substrate effect in terms of a single well-understood quantity, leaving more accurate energy calculations to future research. As our recent work has shown (see Ref. \cite{Kruglov2023CrystalTemperatures}), the most laborious part here is the calculation of the vibrational entropy contribution, since one must use accurate large-scale MD simulations with the pre-trained MLIPs to obtain good results. Fortunately, the correct MLIP for the Mo-S/Al$_2$O$_3$ system has already been trained in this study, allowing us to study the entropy effects in the future.

We also compared our results with the recent work on the study of stable 2D Mo-S structures \cite{Sukhanova20222D-Mo3S4MoS2}, where the standard DFT approach was used to perform the local relaxations and energy evaluations. However the set of stable structures in this work is completely different from what we have obtained, most of these stable structures (except Mo$_5$S and Mo$_5$S$_4$) were successfully found in our simulations. In Supplementary Fig. 8, we show the convex hull with the evolutionary search results without the substrate, where the structures from Ref. \cite{Sukhanova20222D-Mo3S4MoS2} were added for stability comparison. All these reference structures appear to be above the convex hull, indicating their thermodynamic instability. There could be several reasons for such a discrepancy. First, there is no robust and trustworthy way to determine whether the evolutionary algorithm has successfully found the global minimum. Compared to Ref. \cite{Sukhanova20222D-Mo3S4MoS2}, we performed the relaxation of the structures at finite temperature. This usually leads to a smoother potential energy landscape and eliminates some of the local minima that exist in the zero temperature DFT relaxation \cite{Kruglov2023CrystalTemperatures} and could cause the evolutionary search to get stuck. Another related reason is the dynamical instability of some of the thermodynamically stable structures found in ref. \cite{Sukhanova20222D-Mo3S4MoS2}, such as Mo$_5$S and Mo$_5$S$_4$. Due to the finite temperature relaxation, the structures located at saddle points on the potential energy surface could undergo the structural transformation and thus never be found during the evolutionary search. 

\section{Electronic and phonon properties of the stable 2D Mo-S crystals} \label{sec:bandstructures_and_phonons}

Another interesting part besides the stability study is the determination of the electronic properties of the new structures. In the Supplementary Fig. 9, we show the DFT-calculated electronic band structures of all five Mo-S structures presented earlier. Most of the stable structures found during the calculation actually show metallic behavior, except for the common semiconducting MoS$_2$. The latter has a direct DFT band gap of 1.8 eV, which is in perfect agreement with existing DFT results in the literature \cite{Ryou2016MonolayerTransistors}.  What is special about the MoS$_2$ monolayer is that the DFT band gap value is almost identical to that obtained from photoluminescence (PL) and optical absorption experiments \cite{Splendiani2010EmergingMoS2} (1.85 eV). Although, DFT calculations are known to underestimate the band gap values due to the lack of electronic correlations, here it provides better agreement with the experiment compared to the GW approach (2.8 eV) \cite{Ryou2016MonolayerTransistors}. However, this is mostly due to the strong exciton binding ($\sim$ 1 eV), which overcomes the lack of correlations and can be observed in more accurate first-principles methods (e.g. solving the Bethe-Salpeter equation (BSE)). \cite{Ramasubramaniam2012LargeDichalcogenides, Komsa2012EffectsPrinciples, Shi2013QuasiparticleWS2}. 

Since the other stable structures are rich in molybdenum, the electron density is mostly concentrated on molybdenum atoms, which determine the metallic behavior of the corresponding structures. This property can be extremely useful in terms of fabrication of low-dimensional electronic and optical devices, since Mo$_3$S$_2$ and Mo$_5$S$_3$ layers can actually coexist with MoS$_2$ according to the convex hull (as shown in Fig. \ref{fig:ea_results}). This fact allows the creation of 2D metal-semiconductor lateral heterostructures, where the metallic phase can act as an electrode \cite{Sukhanova20222D-Mo3S4MoS2}. 

In addition to the thermodynamic stability of the structures, it's crucial to perform the dynamic stability analysis to better understand the way the material behaves in real applications. Moreover, it is also quite interesting to see if the presence of substrate can change the dynamical stability of the material. We first performed the calculation of the phonon band structures for all the new free-standing 2D crystals from Supplementary Table 2 using the trained MLIP for energy and force evaluation (see Supplementary Figure 10).  All structures from the list except Mo$_3$S$_2$ show no imaginary modes in the phonon spectrum, indicating their dynamical stability. Mo$_3$S$_2$ does indeed have a minor Y-point instability, assosiated by the out-of-plane vibration mode of the 2D layer. All phonon band structures in Supplementary Fig. 10 were calculated using 3x3x1 supercells in a quasi-harmonic approximation using a finite-displacements method.  

To ensure that our MLIP is accurate enough to predict the dynamical stability of the structures, we performed a comparison of DFT and MTP data using 2x2x1 supercells (see Supplementary Fig. 11). First, we noticed that the phonon band structure of almost all crystals changed significantly when using a smaller supercell. In particular, this affected the emergence of $\Gamma$-point instabilities related to out-of-plane modes in both MTP and DFT results while using the smaller supercell. Even though the MTP and DFT bands are not perfectly aligned, most of the instabilities are properly captured and finally disappeared after using a larger supercell. Therefore, it's possible to use the MTP predictions to evaluate both the dynamic and thermodynamic stability of the 2D crystals.

\section{Effect of substrate on phonon properties} \label{sec:phonons_substrate_effect}

\begin{figure}[h]
    \centering
    \includegraphics[width=0.5\textwidth]{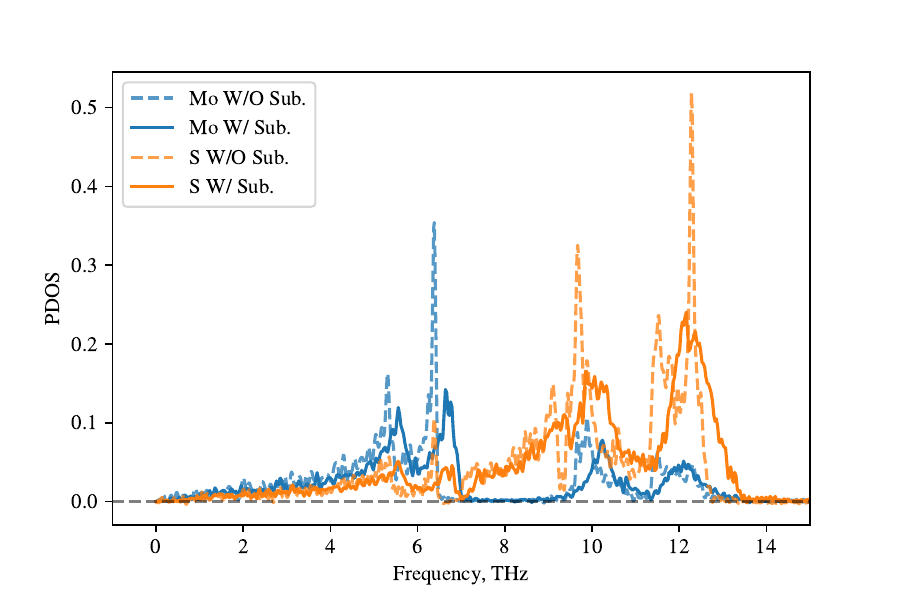}
    \caption{Phonon density of states of the 2D MoS$_2$ layer with (solid lines) and without (dashed lines) substrate. The partial contributions of molybdenum and sulfur atoms are shown in blue and orange, respectively. The effect of the substrate leads to a broadening of the DOS peaks and a shift in their positions.}
    \label{fig:phonon_dos}
\end{figure}

The most interesting part, however, is to find the effect of the substrate on the dynamical stability of the structures. However, this requires the use of a different method to calculate the phonon properties, since the finite displacement method requires the calculation and solution of the eigenproblem of a $3N \times 3N$ dynamical matrix, which is not feasible due to the large size of the systems connected with the substrate. To demonstrate the effect of the substrate on the phonon properties, we calculated the phonon DOS for the MoS$_2$ structure with and without the substrate as the Fourier transform of the velocity autocorrelation function (see section \ref{sec:ph_details} for details). This method not only allows to compute the phonon DOS for large systems, but also reveals a full anharmonic picture of the lattice vibrations. Our results are shown in Fig. \ref{fig:phonon_dos}.  

The interaction with the substrate leads to energy transfer between the 2D layer and the substrate itself, resulting in both shifting and broadening of the DOS peaks. This can affect both the specific heat capacity of the layer and its thermal conductivity, allowing one to design the desired vibrational patterns by choosing a suitable substrate. We also note that similar effects can sometimes be observed when the lattice vibrations are highly anharmonic. Here, however, both the calculations with and without the substrate were performed in a fully anharmonic setup to eliminate this effect. Finally, the effects of anharmonicity on the phonon properties of the 2D MoS$_2$ can be found in the SI. 

\section{Prediction of synthesis conditions}\label{sec:synthesis}

Although the evolutionary search yields a set of thermodynamically stable structures, the experimental synthesis of these structures remains quite complicated and, in fact, hardly related to the computational study we have presented so far. In particular, even the synthesis of 2D MoS$_2$ is represented by at least six different techniques, including CVD growth \cite{Dumcenco2015Large-areaMoS2, Zhu2017CaptureMoS2},  ALD \cite{Song2013Layer-controlledDeposition, Tan2014AtomicFilm, Jin2014NovelSubstrate},  and electron-beam deposition (EBD) \cite{Kong2013SynthesisLayers, Laskar2013LargeMoS2, Lee2014SynthesisPrecursor}. A comprehensive review of 2D MoS$_2$ growth methods can be found in Ref. \cite{Liu2015CVDMaterials}. 

Nevertheless, one of the key properties of the convex hull phase diagrams (Fig. \ref{fig:ea_results}) is a direct relationship between the slope of the convex hull sections and the value of the chemical potential of the components at which they can be stabilized. To further investigate the stability of the structures and to relate them to experimental synthesis conditions, it is necessary to relate the values of the chemical potentials to the parameters that can affect them in each specific experimental setup. In this case, a key quantity that determines the stability of the structure is its Gibbs energy free of formation $G_f$:

\begin{equation}\label{eq:gibbs_formation_energy}
    G_f(P, T) = \frac{1}{N}(G(P,T) - G_{sub}(P,T) - \sum_i n_i \mu_i(P, T))
\end{equation}
where $G(P,T)$ is a Gibbs free energy of a structure joined with the substrate, calculated at a given pressure $P$ and temperature $T$, $G_{sub}(P, T)$ is the Gibbs free energy of a clean substrate,  $n_i, \mu_i(P,T)$ are number of atoms and chemical potential of atomic type $i$ , and $N = \sum_i n_i$ is a total number of atoms in the 2D layer. For each value of $\mu_i$, the most stable structure is then given by a minimum value of the Gibbs free energy of formation among the set of considered structures. For simplicity, we consider only the potential energy contribution to the Gibbs free energies of the structures and the substrate. 

\begin{figure}[htbp]
    \centering
    \includegraphics[width=\textwidth]{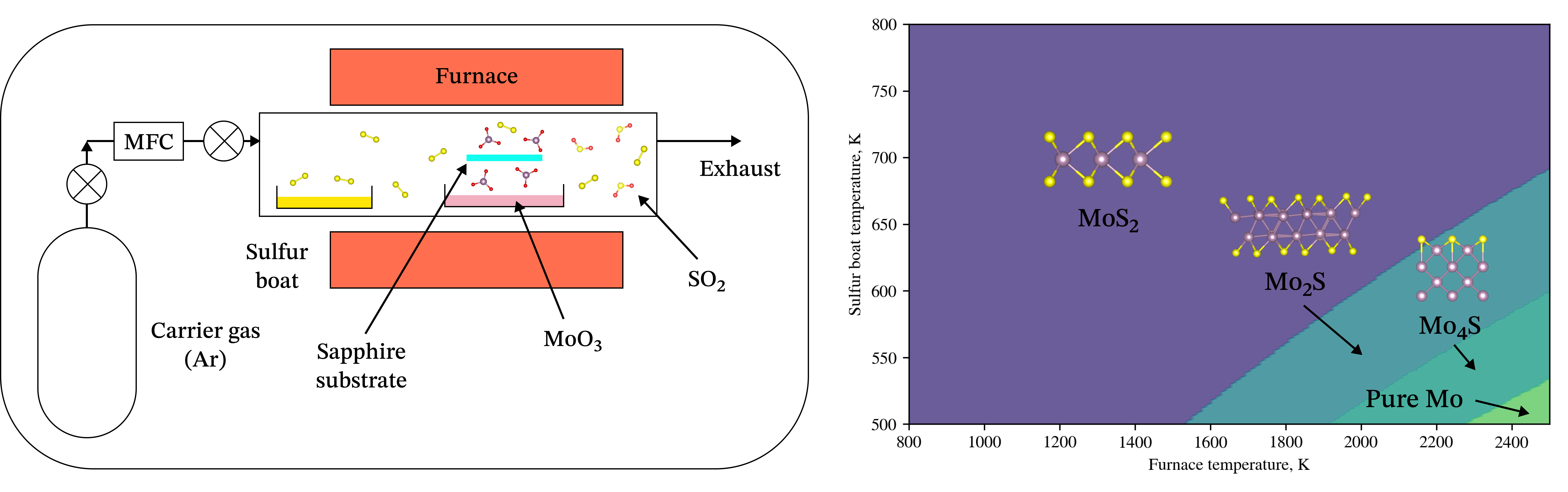}
    \caption{Schematic representation of the CVD synthesis of MoS$_2$ using sulfur and MoO$_3$ precursors (left). In an argon atmosphere, sulfur is vaporized from a sulfur boat and transferred to a furnace in a quartz tube with a sapphire substrate and MoO$_3$ precursor. As the precursors react, MoS$_2$ layers grow on the substrate, and by-products (such as S and SO$_2$ molecules) are removed from the reaction chamber. By controlling the temperature of the furnace ($T_F$) and the sulfur boat ($T_{SB}$), it is possible to alter the chemical potentials of sulfur and molybdenum. Phase diagram of the 2D Mo-S system from the evolutionary search (right), plotted in ($T_F, T_{SB}$) coordinates. At the experimental conditions of MoS$_2$ synthesis ($T_{SB} \sim 800$ K and $T_{F} \sim 1000$ K) \cite{Dumcenco2015Large-areaMoS2}, MoS$_2$ is predicted to be stable. By increasing the $T_F$ and decreasing the $T_{SB}$, it is possible to stabilize other Mo-S phases found in this work.}
    \label{fig:stability_map}
\end{figure}

We used an experimental setup for the CVD synthesis of 2D MoS$_2$ from Ref. \cite{Dumcenco2015Large-areaMoS2} to demonstrate how the synthesis conditions can be predicted from the \textit{ab initio} thermodynamics approach \cite{Reuter2016AbCatalysis}. In this setup, sulfur is vaporized from a sulfur boat and transferred to a furnace in a quartz tube with a sapphire substrate and MoO$_3$ precursor, leading to the formation of MoS$_2$ layers (see left panel in Figure \ref{fig:stability_map}). The vaporization rate of sulfur (controlled by the temperature of the sulfur boat), the temperature of the furnace, and the carrier gas flow rate mostly determine the pressure and temperature of the reaction components. Therefore, we linked the values of the chemical potentials of molybdenum and sulfur to these quantities to find out the stability regions of the structures presented on the convex hull (Fig. \ref{fig:ea_results}). First, we assumed that the partial pressure of sulfur is determined by a certain fraction of its saturated vapor pressure at the temperature of the sulfur boat $T_{SB}$, and sulfur is transferred to the reaction chamber without loss. Additionally, even though sulfur vapor can exhibit different molecular configurations (from S$_2$ to S$_8$), we nevertheless considered only S$_2$ molecules, since their concentration is predominant at $T > 1000$ K \cite{Rau1973ThermodynamicsVapour}. Similarly, the partial pressure of MoO$_3$ molecules is determined by a fraction its saturated vapor pressure at the furnace temperature $T_F$. Finally, we assumed that in the thermodynamic limit, the reaction between MoO$_3$ and sulfur leads to the formation of molybdenum and SO$_2$ gas:

\begin{equation}
    \mathrm{S_2(g) + MoO_3 (g) \rightarrow Mo (s) + SO_2(g)}.
\end{equation}

Therefore, by using the ideal gas approximation for sulfur vapor, MoO$_3$ vapor and SO$_2$ gas, we derived the chemical potentials of molybdenum and sulfur, and used them to analyze the stability of the structures from EA search. A detailed explanation of the underlying calculations, as well as the theoretical background of the \textit{ab initio} thermodynamics is presented in the SI. 

Our results are shown in the right panel of Figure \ref{fig:stability_map}. For simplicity, we assumed that the flow rate of the carrier gas is constant (the reliability of this assumption is discussed further in the text), and only the temperatures of the furnace $T_F$ and the sulfur boat $T_{SB}$ affect the chemical potentials of molybdenum and sulfur. As expected, MoS$_2$ remains thermodynamically stable in a wide range of temperatures, including the experimental synthesis conditions ($T_{SB} \sim 800$ K and $T_{F} \sim 1000$ K)  \cite{Dumcenco2015Large-areaMoS2}. Both decreasing the temperature of the sulfur boat (and thus decreasing the sulfur vaporization rate) and increase in the furnace temperature (promoting the MoO$_3$ evaporation) lead to stabilization of the new structures with higher molybdenum content. 

While interpreting these results, it is important to keep in mind that the predictions are made in a purely thermodynamic limit and do not take into account the kinetics of the process. For this reason, altering the carrier gas flow rate goes beyond the scope of our model. The specific values of the temperatures $T_{F}$  and $T_{SB}$ are therefore descriptive in their nature, and should serve as a guide to the actual experimental procedure rather than a solid synthesis recipe. Nevertheless, the universality of the proposed approach allows the construction of phase diagrams for almost any experimental setup, as soon as the relationship between the synthesis parameters and the chemical potential of the components is established.

\section{Conclusion}\label{sec:conclusions}

In this study, we have developed and demonstrated a comprehensive method for predicting the atomic structure and stability of two-dimensional (2D) materials on arbitrary substrates together with their experimental synthesis conditions. By integrating an evolutionary algorithm with a lattice-matching subroutine and machine learning interatomic potentials (MLIP) for relaxation, we have enabled efficient exploration of a wide range of atomic configurations and chemical compositions. The proposed ASCT algorithm for MLIP training allows the automatic generation of fast and accurate interatomic potentials for arbitrary simulation scenarios, including crystal structure prediction. Our approach was validated on the Mo-S system on a $c$-cut sapphire (Al$_2$O$_3$) substrate, demonstrating its effectiveness in identifying stable 2D crystal structures and evaluating their electronic and phonon properties. The evolutionary search revealed several new stable configurations: $Pmma$ Mo$_3$S$_2$, $P\bar{1}$ Mo$_2$S, $P2_1m$ Mo$_5$S$_3$, and $P4mm$ Mo$_4$S, where the Mo$_4$S structure is specifically stabilized by interaction with the substrate.  These structures extend the known landscape of 2D materials and open new possibilities for practical applications. Electronic band structure calculations showed that all the new structures exhibit metallic behavior. This suggests the potential to create 2D metal-semiconductor lateral heterostructures, where the metallic phases could serve as electrodes in electronic devices. Phonon calculations showed that all but one of the identified structures are dynamically stable. The Mo$_3$S$_2$ structure showed a small Y-point instability, associated with the out-of-plane vibration modes. Interaction with the substrate was found to alter the phonon density of states, potentially allowing modulation of the thermal and mechanical properties of the 2D materials. Based on the \textit{ab initio} thermodynamics approach, we demonstrated the phase diagram of the predicted 2D Mo-S structures in the parameter space of the CVD experimental setup.  Our results highlight the importance of including substrate effects in computational studies to accurately predict the stability and properties of 2D materials. The trained MLIP not only facilitates accurate predictions, but also significantly reduces computational costs, making it feasible to study large systems that are otherwise intractable with conventional DFT methods. Overall, our method opens new horizons in 2D materials discovery, allowing to find new substrate-stabilized phases even in well known and thoroughly studied systems, obtain the configurations for characterizing the substrate effect on various properties of 2D materials, and predict the possible conditions for their synthesis to design the next generation of electronic and optoelectronic devices. Future work will focus on further refining the MLIP training process and extending the application of this method to other material systems and substrates, aligning better the computational approaches used with real experimental setups, and experimental verification of newly discovered 2D materials.

\section{Methods}\label{sec:methods}

\subsection{MLIP training details for 2D Mo-S/Al$_2$O$_3$ system}\label{sec:mlip}

\subsubsection{2D Mo-S system}\label{sec:mlip_mo_s}

The ML potential for the Mo-S/Al$_2$O$_3$ system was trained within the ASCT framework in three steps. In the first stage, random 2D structures in the Mo-S system with 2 to 16 atoms in the unit cell were generated using the symmetry-based random structure generator from a PyXtal package \cite{Fredericks2021PyXtal:Analysis}. These structures were then used to initialize the sampling in 20 parallel MD simulations. Each MD run began with a replication of the unit cell in the in-plane directions to obtain the structure with approximately 64 atoms per unit cell. The following steps were then performed for each structure:

\begin{enumerate}
    \item A conjugate gradient relaxation of the atomic positions with a fixed cell and a convergence criterion of 10$^{-10}$ eV. 
    \item NPT annealing at T = 10 K and P = 1 bar for 10 ps
    \item NPT heating from 10 K to 300 K at P = 1 bar for 50 ps
    \item NPT anneal at T = 300 K and P = 1 bar for 50 ps
\end{enumerate}

A time step of 1 fs was used for all simulations. The MTP extrapolation level thresholds for sampling and run interruption were set to 3 and 10, respectively. The cutoff radius for the local atomic environment representation was 5 \AA. The complexity of the model in terms of the size of the basis set was fixed by selecting the 24g.mtp initial potential file. Each iteration of training was performed with weights of 10, 0.01, 0.001 for energies, forces, and stresses contributions in the loss function, while the weight scaling for energies and forces was 2 and 1, respectively. The ASCT convergence criterion was set to 25 iterations. Finally, the ML potential for the 2D Mo-S system was obtained after 41 iterations of the ASCT and the training set contained 727 structures. 

\subsubsection{Al$_2$O$_3$ surface slab}\label{sec:mlip_al2_o3}

In the second stage, the \textit{c}-cut surface slab of $\alpha$-Al$_2$O$_3$ with 48 Al and 72 O atoms and a thickness of 10 \AA \ was used to initialize the sampling and to train a separate interatomic potential. The ASCT routine and MTP training parameters were almost identical to those used in the case of a Mo-S system, except for the replication of the unit cell and a lower convergence criterion of 15 iterations. Thus, the potential was obtained in 27 iterations of ASCT with a total of 1140 structures in the training set.

\subsubsection{2D Mo-S/Al$_2$O$_3$ }\label{sec:mlip_mo_s_al2_o3}

In the final, third step, the training sets for the Mo-S and Al$_2$O$_3$ systems were merged to train the initial version of the desired Mo-S/Al$_2$O$_3$ potential. This time the MD sampling part of ASCT was initialized with the random Mo-S structures containing up to 16 atoms in the unit cell, stacked with a surface plate of Al$_2$O$_3$ using a lattice matching algorithm (see section \ref{sec:lattice_matching}). To avoid abnormally large structures, we did not apply lattice multiplication in this step and just fit the generated random structure into the fixed unit cell of the Al$_2$O$_3$ surface slab. Despite the lattice mismatch strain induced by this operation, these structures are still suitable for training the interatomic potential, especially given their "random" nature inherent in their origin. The rest of the training was done in the same way as in the previous cases.  The convergence criterion of ASCT was 15 iterations, and the training was done in a total of 37 iterations with 3218 structures in the training set. 

\subsection{Details of the \textit{ab initio} calculations}\label{sec:dft}
 
\subsubsection{Calculation of data for MLIP training}

Training data on energies, interatomic forces and stresses of the new configurations within the ASCT workflow were obtained at the density functional theory (DFT) level using the VASP package \cite{Kresse1996EfficiencySet, Kresse1996EfficientSet}. The convergence criteria for an electronic self-consistent cycle was 10$^{-6}$ eV, and the cutoff energy for a plane-wave basis set was 600 eV. We used a Gaussian smearing scheme with a width of 0.05 eV to represent the band occupancy. The first Brillouin zone was represented by a $\Gamma$-centered uniform lattice with a density of $2\pi \cdot 0.03$ Å. The behavior of the core electrons and their interaction with the valence electrons was described within Projector-Augmented Wave (PAW) pseudopotentials \cite{Kresse1999FromMethod} with 12 [4p5s4d], 6 [s2p4], 3 [s2p1], and 6 [s2p4] valence electrons for Mo, S, Al, and O, respectively. The electronic exchange correlation effects were modeled within a generalized gradient approximation (Perdew-Burke-Ernzerhof functional) \cite{Perdew1996GeneralizedSimple}. A vacuum layer of 20 \AA \ was added to all structures along the direction normal to the surface plane to avoid surface interactions due to periodic boundary conditions. 

\subsubsection{Band structure calculations}

The electronic band structure calculations were performed at the DFT level using a \textit{RelaxBandStructureMaker} utility implemented in the \textit{atomate2} package \cite{ganose_atomate2_2024} and the VASP backend. All crystal structures (i.e. their positions and cell parameters) were first relaxed until all interatomic forces were less than $2 \cdot 10^{-2}$ eV/\AA. Next, the standard SCF calculation was performed to obtain the Kohn-Sham orbitals on a coarse k-point grid with a spacing of 0.3 \AA$^{-1}$. Finally, a band structure calculation was performed for a specific k-path generated by the SeeK-path utility \cite{Hinuma2017BandCrystallography} based on crystal symmetry. We used the same set of PAW potentials as for MLIP training and evaluation, but chose a slightly higher cutoff of 680 eV to accurately describe the surface effects of the electrons. We also switched to the PBEsol functional \cite{Perdew2008RestoringSurfaces} to describe the exchange-correlation effects of the electrons, as this usually results in more accurate electronic band structures. Partial occupancies of the orbitals were set using a Gaussian smearing method with a width of 0.01 eV. 

\subsubsection{Phonon properties calculations}\label{sec:ph_details}

For the free-standing 2D crystals, phonon band structures and densities of states were calculated in a quasi-harmonic approximation with a \textit{Phonopy} package \cite{Togo2015FirstScience, Togo2022First-principlesPhono3py, Togo2023ImplementationPhono3py}. We used both VASP- and MTP-based relaxation and force constant evaluation to evaluate the accuracy of the trained ML potential. First, the positions and cell parameters of all crystals were relaxed until the interatomic forces were less than $5 \cdot 10^{-4}$ eV/\AA. Next, the set of symmetric nonequivalent finite displacements was generated using \textit{Phonopy} utilities, and the resulting interatomic forces were calculated to evaluate the dynamical matrix. This matrix was finally diagonalized to obtain a set of eigenfrequencies for the corresponding set of wavevectors. Compared to previous DFT calculations, we used a slightly denser k-point grid with a spacing of 0.2 \AA$^{-1}$, but a larger electronic smearing of 0.05 eV, as this helped to obtain better converged results. 

For the systems connected to the substrate, we used a different approach to calculate phonon density of states based on the Fourier transform of the velocity autocorrelation function (VACF) (Eq. \eqref{eq:pdos}). 

\begin{equation}
    \label{eq:pdos}
    g(\nu) = 4 \int_{0}^{\infty} cos(2\pi \nu t) \frac{\langle \overline{\nu(0) \nu(t)} \rangle}{\langle \overline{\nu(0)^2} \rangle}
\end{equation}

VACF was calculated during NVE molecular dynamics simulation (with trained MTP potential) for each studied system. It allows one to calculate the phonon DOS in the full anharmonic picture and is generally more accessible for large systems than a finite displacement method.

We also used the same approach to calculate the phonon DOS for the freestanding 2D layers, in order to identify the effect of the substrate and the potential influence of the anharmonicity.

\subsection{Details of the evolutionary search}\label{sec:ea_details}

The evolutionary search for the new stable structures in a 2D Mo-S/Al$_2$O$_3$ system was performed with a modified version of the USPEX code.  During the search, the structures were allowed to have from 4 to 16 atoms of variable Mo-S composition in the unit cell, while their thickness was constrained to 6 \AA. Each generation of the evolutionary search consisted of 120 structures, except for the first one, which had 180 structures. The first generation was generated with a symmetry-based random structure generator, while structures in all subsequent generations were generated with a heredity (40 \%) and transmutation (30 \%) evolutionary operator using the subset of the best structures from the previous generation. The heredity operator has been adapted to 2D crystals from its 3D analog in the USPEX code \cite{Oganov2006CrystalApplications, Oganov2011HowWhy}. It first slices two parent structures along a random direction and then alternately combines the slices to create a new structure. The transmutation operator uses only one parent structure and creates a new one by randomly assigning new chemical identities to a group of atoms. 
The remaining 30 \% of the structures in each generation were generated randomly. The total number of generations was limited to 150, while the evolutionary search was considered to converge when the list of best structures remained unchanged for 25 consecutive generations. 

Local relaxation of the structures was performed in three steps. In the first step, the generated 2D Mo-S configurations were relaxed and annealed with the pre-trained MTP potential and a LAMMPS package, following the same protocol used to train the MLIP (see section \ref{sec:mlip_mo_s}). Next, the relaxed 2D structures were joined to the Al$_2$O$_3$ substrate using a lattice matching algorithm (see section \ref{sec:lattice_matching}) with an initial gap value of 2.0 \AA, a maximum mismatch criterion of $5 \cdot 10^{-3}$, and a maximum value of the resulting substrate area of 1000 \AA$^2$. This joint structure was again relaxed and annealed in the same manner. In the last step, the resulting configuration of the substrate was used for a single point calculation of the substrate energy for further calculation of the fitness function of the structures. Finally, as the relaxation is done, the fitness function of the structures was calculated in terms of the energy over the composition convex hull. To do this, we first calculate the formation energy $\Delta E_f$ of each structure $\Delta E_{f} = \frac{1}{N} (E - \sum_i n_i E_i)$ , where $E$ is the energy of the structure, $n_i$, $E_i$ are the number of atoms and energies of each pure component, and $N = \sum_i n_i$ is the total number of atoms in the structure. The energy of the structure is either equal to the total energy from the DFT calculation $E_{tot}$, or to the difference between the total energy and the clean substrate energy $E_{sub}$: $E=E_{tot} - E_{sub}$, depending on whether the free-standing 2D layers or those joined with the substrate are considered. Finally, we build the convex hull in the ($\Delta E_f$ - composition) space and evaluate the energy over the convex hull to determine the stability of the structure.

\subsection{Lattice-matching algorithm}\label{sec:lattice_matching}

To obtain the appropriate supercell representation of the 2D material on top of the substrate, we adapted the Zur-McGill lattice matching algorithm \cite{Zur1984LatticeHeteroepitaxy} implemented in the PyMatGen package \cite{Ong2013PythonAnalysis}.  The algorithm essentially builds a set of the supercell matrices for both substrate and 2D layer unit cells to eventually generate two roughly equal supercells with a desired value of the estimated mismatch, which is based on a ratio of the areas of the resulting lattices. We also constrain the maximum area of the supercells to obtain a Pareto optimal solution in terms of lattice mismatch and system size. This requires a series of trial runs of the algorithm with different values of the maximum area, followed by an analysis of the resulting lattice mismatch. Usually, large values of the maximum area lead to structures with low mismatch but a large number of atoms. It is therefore necessary to choose appropriate parameters before starting the calculations, depending on the desired level of accuracy and the available computational resources. As the supercells are generated, the algorithm connects the 2D layer to a substrate at a given gap distance.

 \backmatter

\section*{Data availability}
All the data produced in this work is available at the Materials Cloud Archive after publication.

\section*{Code availability}
The code will be merged into the next release of the USPEX code (\href{https://uspex-team.org/}{https://uspex-team.org/})

\section*{Acknowledgments}

I.A.K. thanks grant RSF No. 24-73-10055 for financial support

\section*{Competing interests}
The authors declare no competing interests.

\bibliography{main}

\end{document}

\title[Article Title]{Supplementary Information: Substrate-aware computational design of two-dimensional materials}

\author*[1]{\fnm{Arslan} \sur{Mazitov}}\email{arslan.mazitov@phystech.edu}

\author[1,2]{\fnm{Ivan} \sur{Kruglov}}

\author[1]{\fnm{Alexey V.} \sur{Yanilkin}}

\author[1,2,3]{\fnm{Aleksey V.} \sur{Arsenin}}

\author[2,3]{\fnm{Valentyn S.} \sur{Volkov}}

\author[4]{\fnm{Dmitry G.} \sur{Kvashnin}}

\author[5]{\fnm{Artem R.} \sur{Oganov}}

\author[6,7,8]{\fnm{Kostya S.} \sur{Novoselov}}

\affil[1]{\orgname{Moscow Center for Advanced Studies}, \orgaddress{\street{Kulakova str. 20}, \city{Moscow}, \postcode{123592}, \country{Russian Federation}}}

\affil[2]{\orgdiv{Emerging Technologies Research Center}, \orgname{XPANCEO}, \orgaddress{\street{Internet City, Emmay Tower}, \city{Dubai},  \country{United Arab Emirates}}}

\affil[3]{\orgdiv{Laboratory of Advanced Functional Materials}, \orgname{Yerevan State University}, \city{Yerevan}, \postcode{0025}  \country{Armenia}}

\affil[4]{\orgname{Emanuel Institute of Biochemical Physics}, \orgaddress{ \street{Kosigina st. 4}  \city{Moscow}, \postcode{119334},  \country{Russian Federation}}}

\affil[5]{\orgdiv{Materials Discovery Laboratory}, \orgname{Skolkovo Institute of Science and Technology (Skoltech)}, \orgaddress{\street{Bolshoy Boulevard 30, bld. 1}, \city{Moscow}, \postcode{121205},  \country{Russian Federation}}}

\affil[6]{\orgdiv{National Graphene Institute (NGI)}, \orgname{University of Manchester}, \orgaddress{\city{Manchester}, \postcode{M13 9PL},  \country{UK}}}

\affil[7]{\orgdiv{Department of Materials Science and Engineering}, \orgname{National University of Singapore}, \orgaddress{\city{Singapore}, \postcode{03-09 EA},  \country{Singapore}}}

\affil[8]{\orgdiv{Institute for Functional Intelligent Materials}, \orgname{National University of Singapore}, \orgaddress{\city{Singapore}, \postcode{117544},  \country{Singapore}}}

\maketitle

\section{Performance of the trained MLIP on the training and validation sets }\label{sec:energy_pairplots}

\begin{figure}[H]
    \centering
    \includegraphics[width=\linewidth]{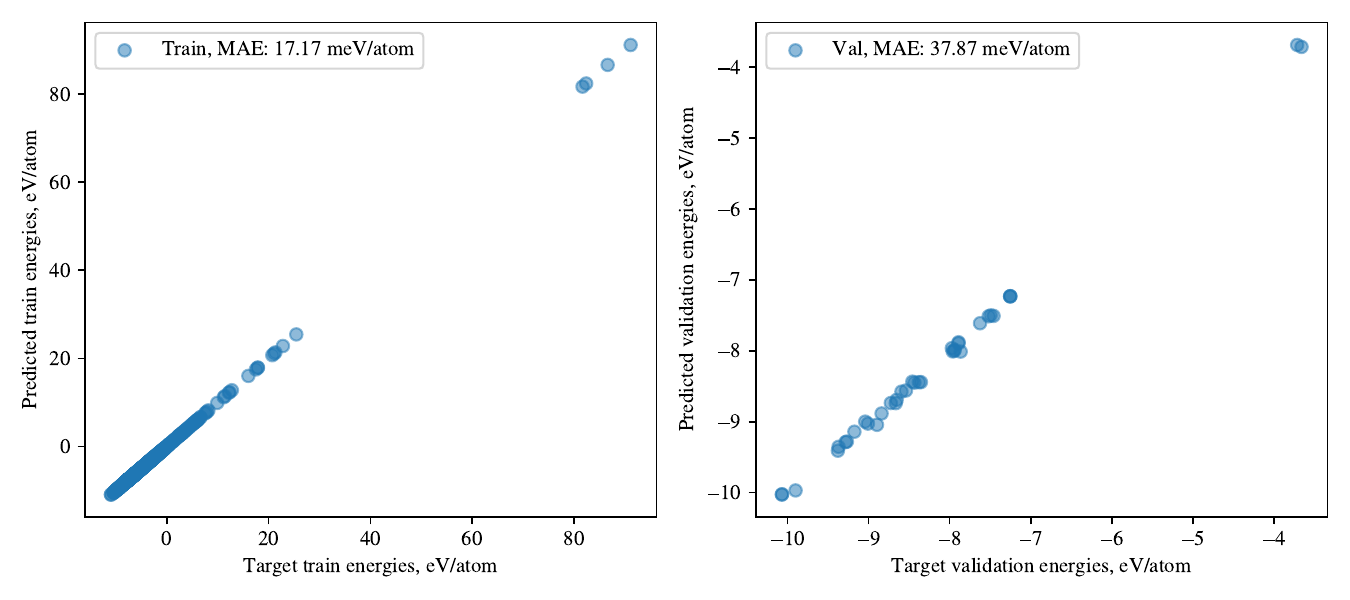}
    \caption{Training (left) and validation (right) energy pair plots computed with the trained MLIP. A set of stable and metastable structures from the EA search used for validation.}
    \label{fig:energy_pairplots}
\end{figure}

\section{Structure and stability data for the stable and metastable phases found in the evolutionary search}\label{sec:ea_structures_lattice_parameters}
\begin{table}
    \centering
    \begin{tabular}{|c|c|c|c|c|c|c|} \hline 
         Composition&  Space group&  $a$&  $b$&  $\alpha$&  $\beta$& $\gamma$\\ \hline 
         MoS$_2$&  $P\bar{6}m2$&  3.182&  3.182&  60&  90& 90\\ \hline 
         Mo$_2$S&  $P\bar{1}$&  5.494&  5.263&  89.11&  94.17& 80.25\\ \hline 
         Mo$_5$S$_3$&  $P2_1m$&  4.430&  8.261&  90.95&  90.32& 90\\ \hline 
         Mo$_3$S$_2$&  $Pmma$&  4.446&  5.255&  91.76&  90.25& 90\\ \hline 
         Mo$_4$S&  $P4mm$&  3.152&  3.152&  89.97&  90.12& 90\\ \hline
    \end{tabular}
    \caption{Lattice parameters of the stable and metastable 2D Mo-S structures found in the EA search. All the parameters were obtained after DFT relaxation of the structures originally predicted in the EA run.}
    \label{tab:ea_structures_lattice_parameters}
\end{table}

\begin{table}[h!]
\centering
 \begin{tabular}{|c  |c  |c  |c  |c |c |c |} 
 \hline
  Comp.& $a_{\mathrm{MTP}}$, \AA& $a_{\mathrm{DFT}}$, \AA&  $E_{f}^{\mathrm{MTP}}$,  W/O S.& $E_{f}^{\mathrm{MTP}}$, W/ S.& $E_{f}^{\mathrm{DFT}}$,  W/O S.&$E_{f}^{\mathrm{DFT}}$,  W/ S.\\
 \hline
 MoS$_2$ & 3.15& 3.18& -1.433&-1.337& -1.354&-1.406\\
 \hline
 Mo$_2$S & 5.62& 5.50&-0.820&-0.748& -0.770&-0.693\\
 \hline
 Mo$_5$S$_3$ & 4.40& 4.43& -0.901& -0.847& -0.850&-0.867\\
 \hline
Mo$_3$S$_2$ & 4.39& 4.45& -0.963& -0.893& -0.919&-0.937\\
\hline
Mo$_4$S &  3.11& 3.15& -0.487& -0.467& -0.454&-0.560\\
\hline
 \end{tabular}
 \caption{Comparison of the lattice parameters and formation energies obtained with DFT and MTP for stable and metastable 2D Mo-S structures. Lattice parameters represent the values obtained after MTP and DFT relaxation respectively. Formation energy values correspond to MTP and DFT results of freestanding 2D layers (without substrate) and those joined with the substrate.}
 \label{tab:structures_summary}
\end{table}

\section{Stable and metastable 2D Mo-S structures joined with the substrate}\label{sec:structures_with_substrate}

\begin{figure}[H]
    \centering
    \includegraphics[width=\linewidth]{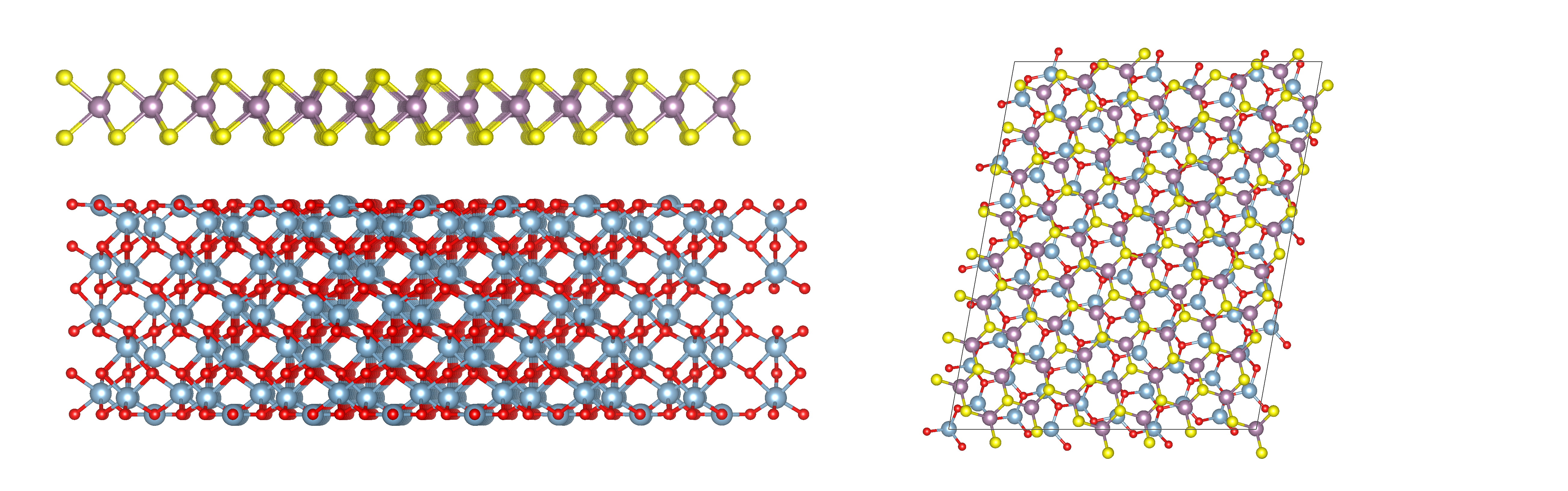}
    \caption{Visualization of the MoS$_2$ structure joined with Al$_2$O$_3$ substrate using a lattice matching algorithm. The purple and yellow spheres represent the molybdenum and sulfur atoms, while the blue and red spheres represent the aluminum and oxygen atoms. The structure are fully relaxed with the MTP potential. All the atoms of the substrate except the top layer are removed from the top view for convenience.}
    \label{fig:MoS2_withEnv}
\end{figure}

\begin{figure}[H]
    \centering
    \includegraphics[width=\linewidth]{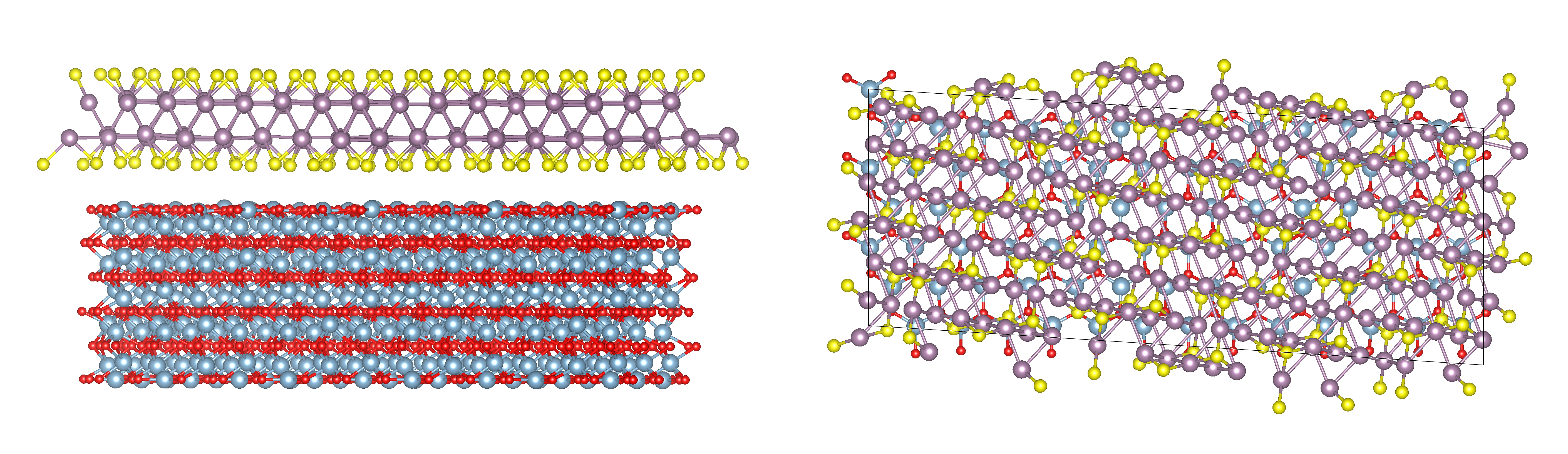}
    \caption{Visualization of the Mo$_2$S structure joined with Al$_2$O$_3$ substrate using a lattice matching algorithm. The purple and yellow spheres represent the molybdenum and sulfur atoms, while the blue and red spheres represent the aluminum and oxygen atoms. The structure are fully relaxed with the MTP potential. All the atoms of the substrate except the top layer are removed from the top view for convenience.}
    \label{fig:Mo2S_withEnv}
\end{figure}

\begin{figure}[H]
    \centering
    \includegraphics[width=\linewidth]{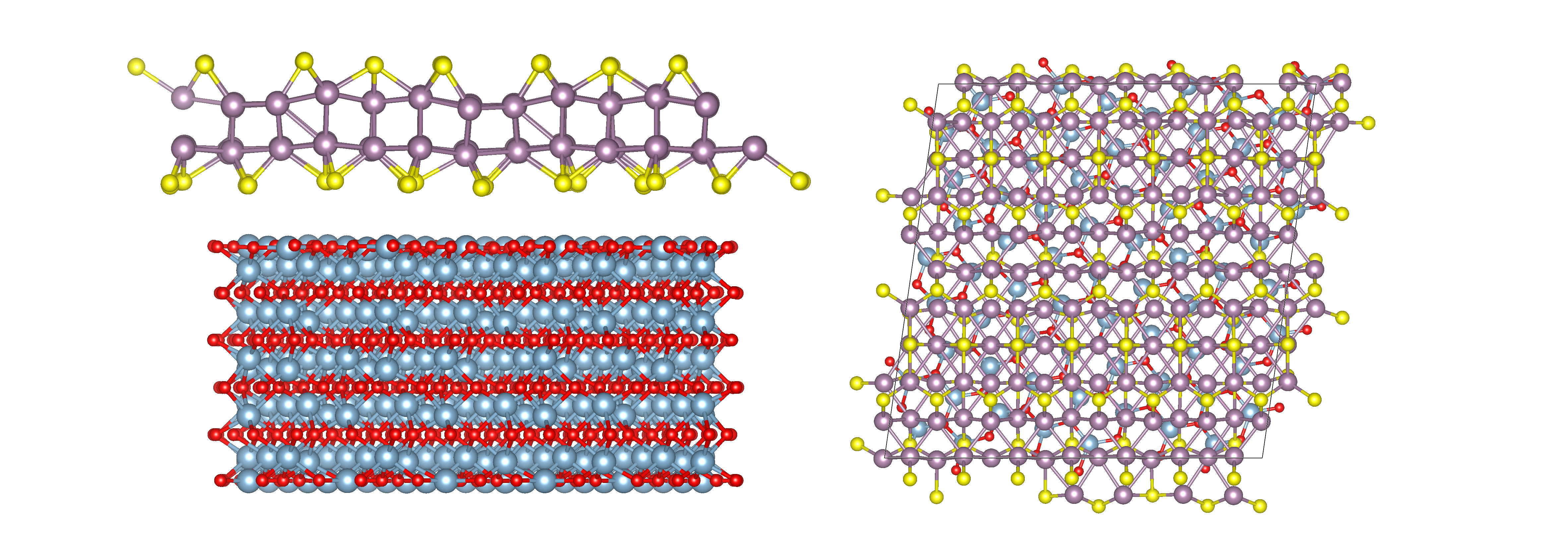}
    \caption{Visualization of the Mo$_5$S$_3$ structure joined with Al$_2$O$_3$ substrate using a lattice matching algorithm. The purple and yellow spheres represent the molybdenum and sulfur atoms, while the blue and red spheres represent the aluminum and oxygen atoms. The structure are fully relaxed with the MTP potential. All the atoms of the substrate except the top layer are removed from the top view for convenience.}
    \label{fig:Mo5S3_withEnv}
\end{figure}

\begin{figure}[H]
    \centering
    \includegraphics[width=\linewidth]{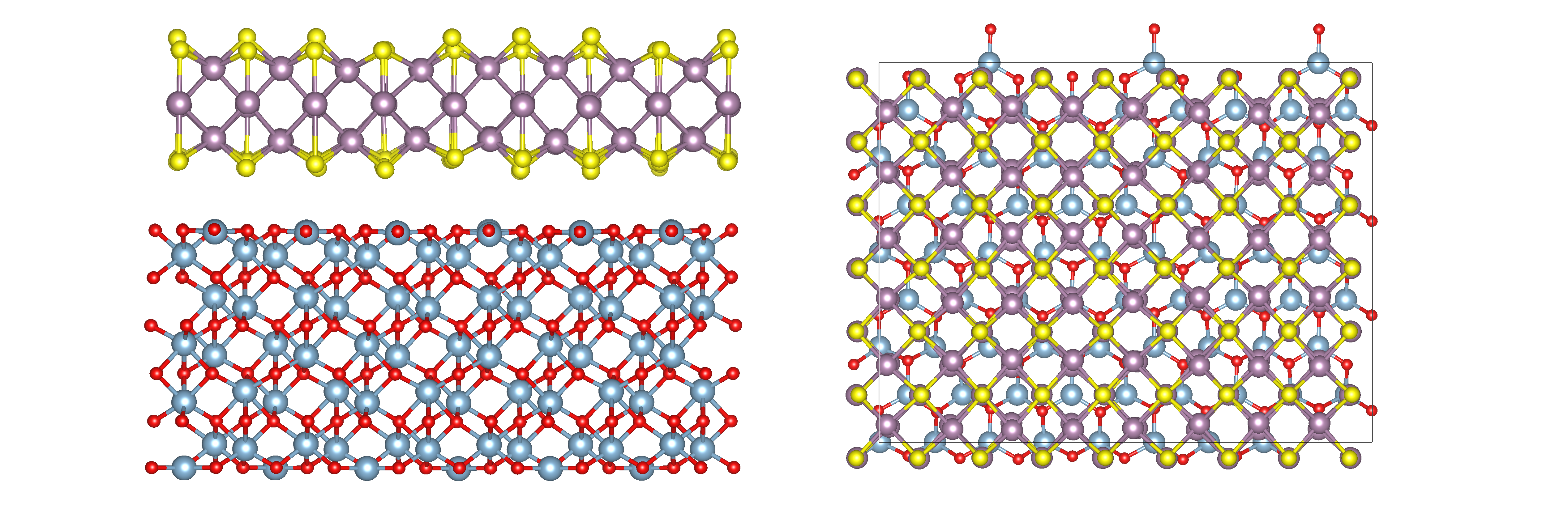}
    \caption{Visualization of the Mo$_3$S$_2$ structure joined with Al$_2$O$_3$ substrate using a lattice matching algorithm. The purple and yellow spheres represent the molybdenum and sulfur atoms, while the blue and red spheres represent the aluminum and oxygen atoms. The structure are fully relaxed with the MTP potential. All the atoms of the substrate except the top layer are removed from the top view for convenience.}
    \label{fig:Mo3S2_withEnv}
\end{figure}

\begin{figure}[H]
    \centering
    \includegraphics[width=\linewidth]{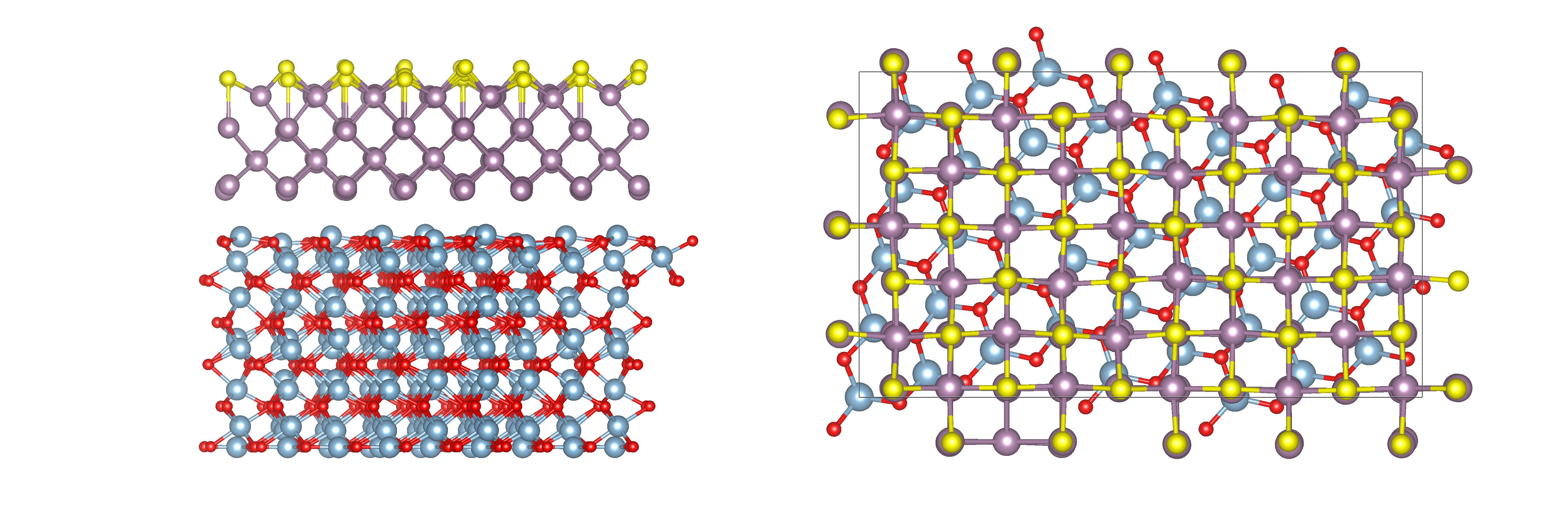}
    \caption{Visualization of the Mo$_4$S structure joined with Al$_2$O$_3$ substrate using a lattice matching algorithm. The purple and yellow spheres represent the molybdenum and sulfur atoms, while the blue and red spheres represent the aluminum and oxygen atoms. The structure are fully relaxed with the MTP potential. All the atoms of the substrate except the top layer are removed from the top view for convenience.}
    \label{fig:Mo4S_withEnv}
\end{figure}

\section{Effect of chemical bonding between 2D layers and substrate}\label{sec:Mo4S_Mo_Al_bonds}

\begin{figure}[H]
    \centering
    \includegraphics[width=\linewidth]{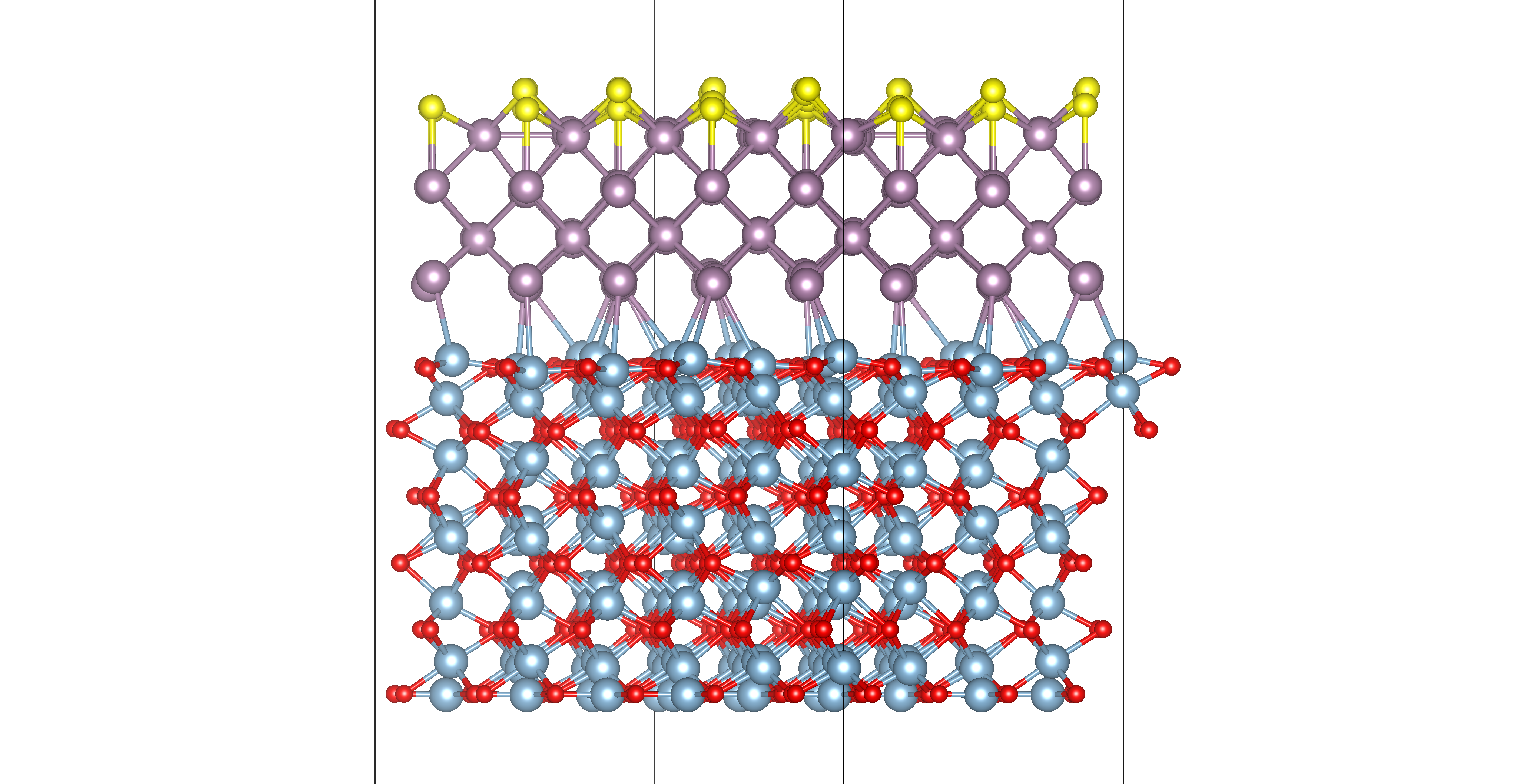}
    \caption{Chemical bonding between Mo and Al atoms in the Mo$_4$S structure joined with Al$_2$O$_3$ substrate. The color coding replicates that used in Supplementary Fig. \ref{fig:Mo4S_withEnv}. Mo-Al bond lengths vary between 3.0 \AA \  and 3.1 \AA, which is close to the average bond lengths in the Mo-Al crystal systems (from 2.8 \AA \ to 3.0 \AA)  \cite{Jain2013Commentary:Innovation}. Chemical bonding with the substrate saturates a certain fraction of the bonds in the 2D layer, stabilizing structures that would not be stable in a free-standing 2D case.}
    \label{fig:Mo4S_Mo_Al_bonds}
\end{figure}

\section{Stable 2D Mo-S structures from the Reference paper}\label{sec:ea_results_with_ref}
\begin{figure}[H]
    \centering
    \includegraphics[width=\linewidth]{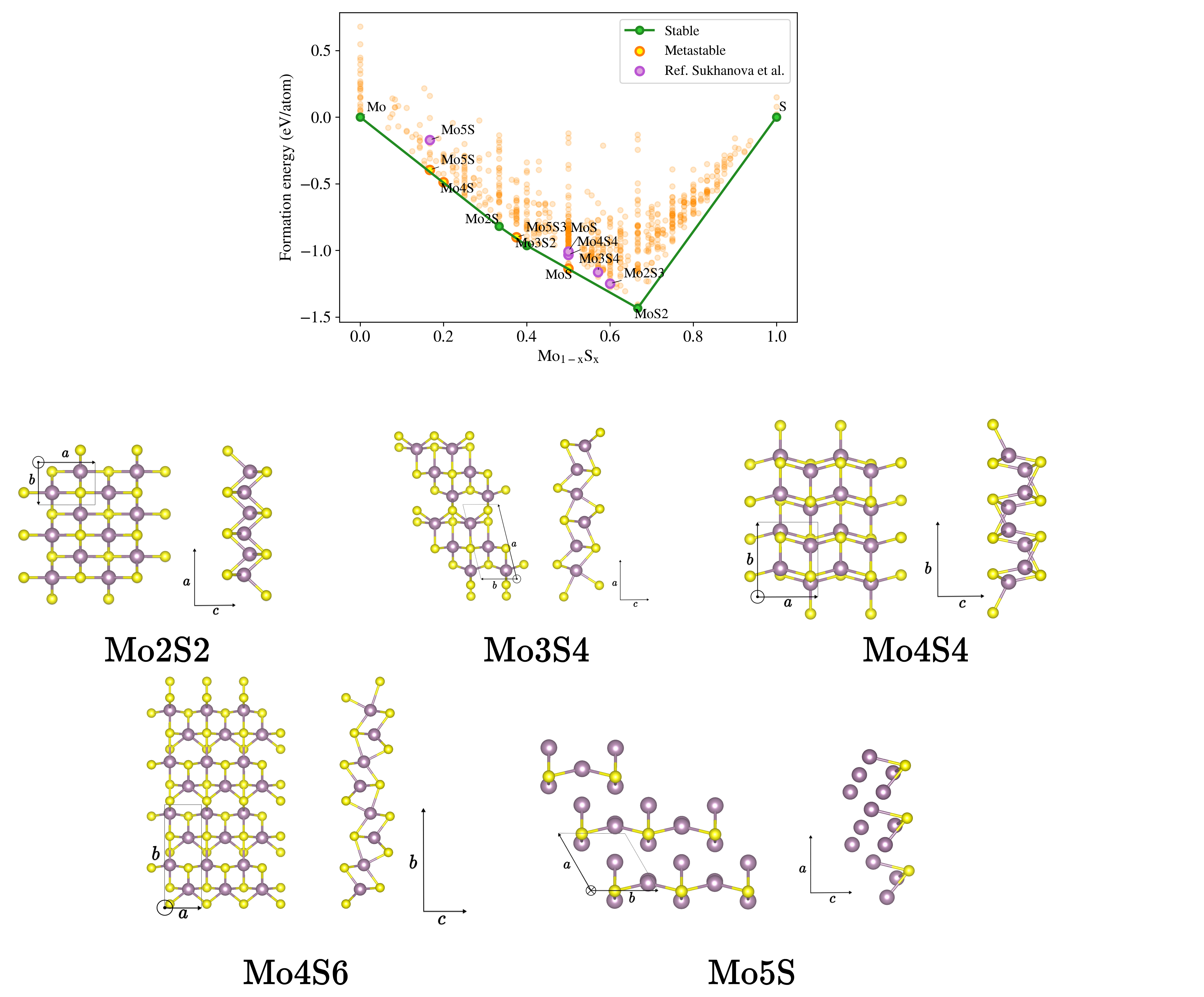}
    \caption{Results of the evolutionary search in the Mo-S @ Al$_2$O$_3$ system with all stable structures from the Ref. \cite{Sukhanova20222D-Mo3S4MoS2} (shown as purple circles). The color coding is made consistent with the Figures in the main text. Most of the structures, except Mo$_5$S and Mo$_5$S$_4$ were successfully found during the evolutionary search, and appear to be above the convex hull, indicating their thermodynamic instability. All the reference structures and stables structures on the convex hull were preliminary relaxed in DFT. The closet analog of the Mo$_5$S is shown instead in the list of crystal structures.}
    \label{fig:ea_results_with_ref}
\end{figure}
\section{Electronic band structures of the stable 2D Mo-S structures}\label{sec:bandstructures}
\begin{figure*}[htb]
    \centering
    \begin{subfigure}[h!]{0.32\textwidth}
        \centering
        \includegraphics[width=\textwidth]{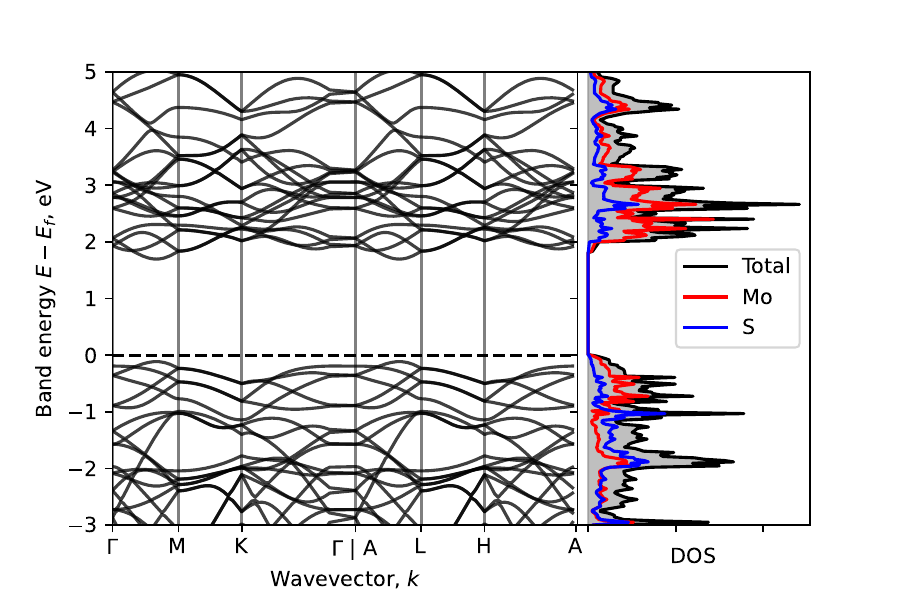}
        \caption{MoS$_2$}
    \end{subfigure}
    \begin{subfigure}[h!]{0.32\textwidth}
        \centering
        \includegraphics[width=\textwidth]{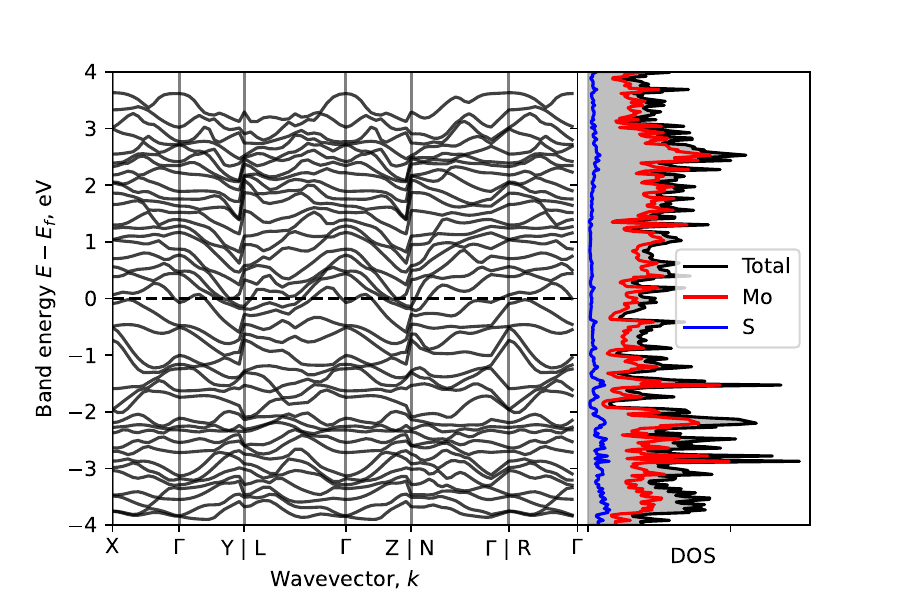}
        \caption{Mo$_2$S  }
    \end{subfigure}
      \begin{subfigure}[h!]{0.32\textwidth}
        \centering
        \includegraphics[width=\textwidth]{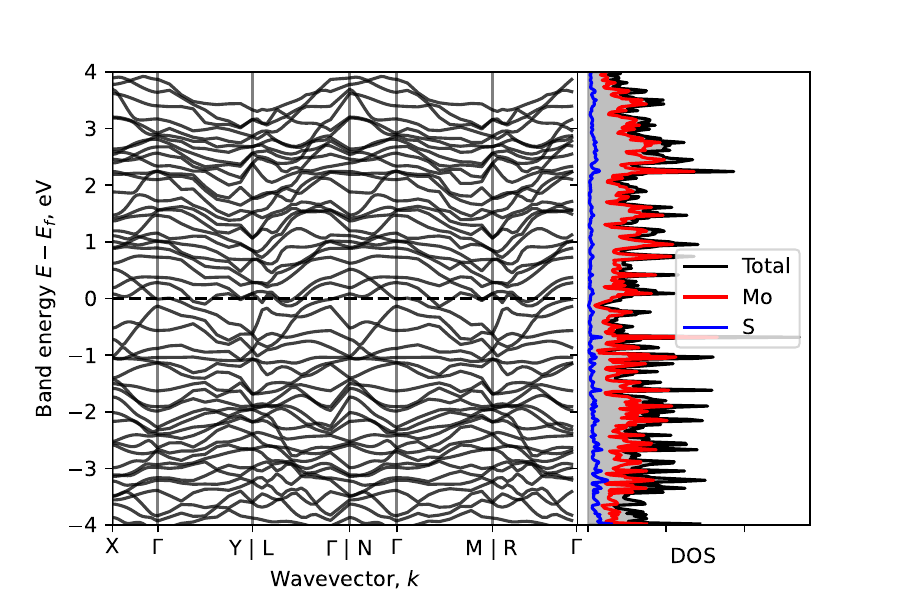}
        \caption{Mo$_5$S$_3$  }
    \end{subfigure}

    \begin{subfigure}[h!]{0.32\textwidth}
        \centering
        \includegraphics[width=\textwidth]{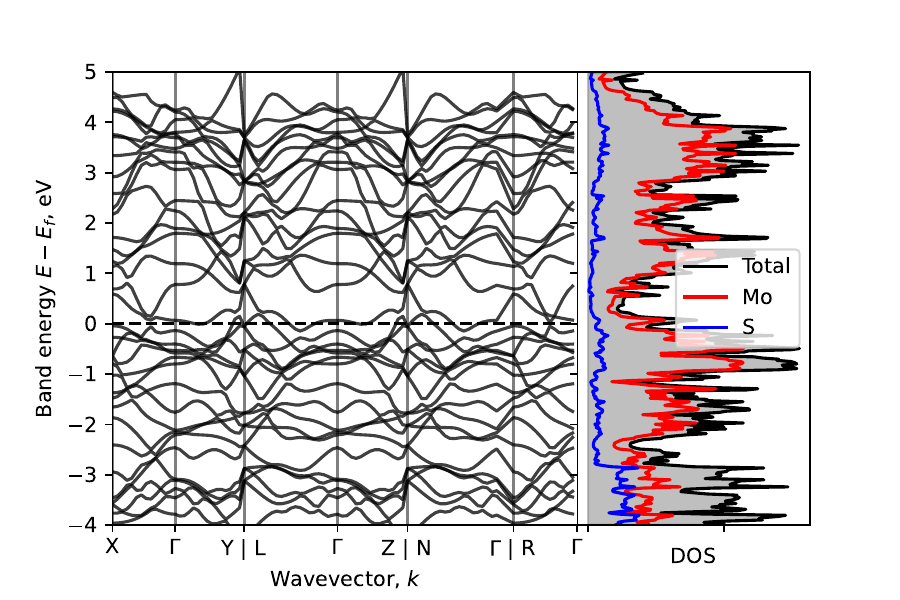}
        \caption{Mo$_3$S$_2$  }
    \end{subfigure}
     \begin{subfigure}[h!]{0.32\textwidth}
        \centering
        \includegraphics[width=\textwidth]{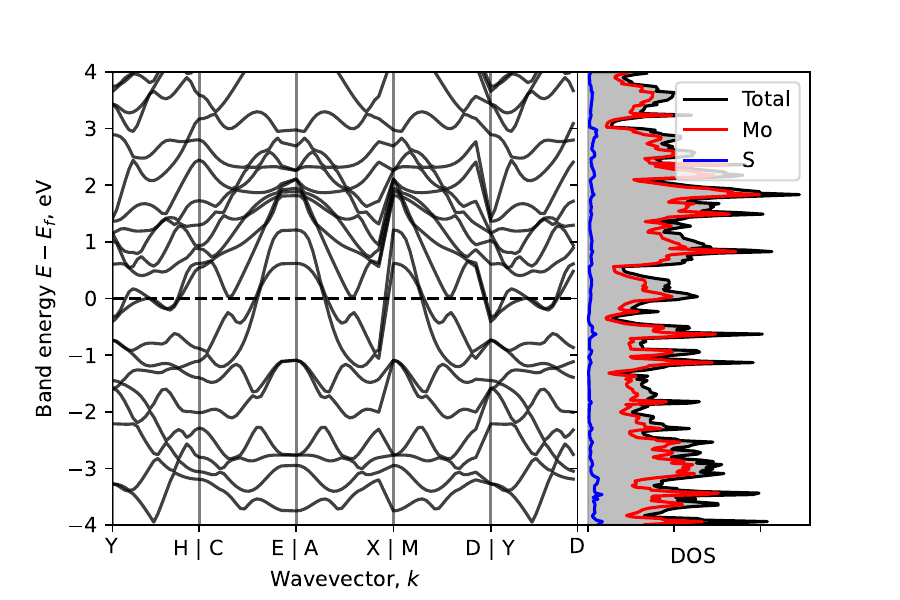}
        \caption{Mo$_4$S}
    \end{subfigure}
    \caption{Electronic band structures of the stable freestanding 2D crystals found during the evolutionary search. All structures except 2D MoS$_2$ demonstrate metallic behavior with electron density mostly concentrated on molybdenum atoms.}
        \label{fig:bs}
\end{figure*}

\section{Phonon band structures of the stable and metastable 2D Mo-S structures}\label{sec:phonon_bs}

\begin{figure*}[htb]
    \centering
    \begin{subfigure}[h!]{0.49\textwidth}
        \centering
        \includegraphics[width=\textwidth]{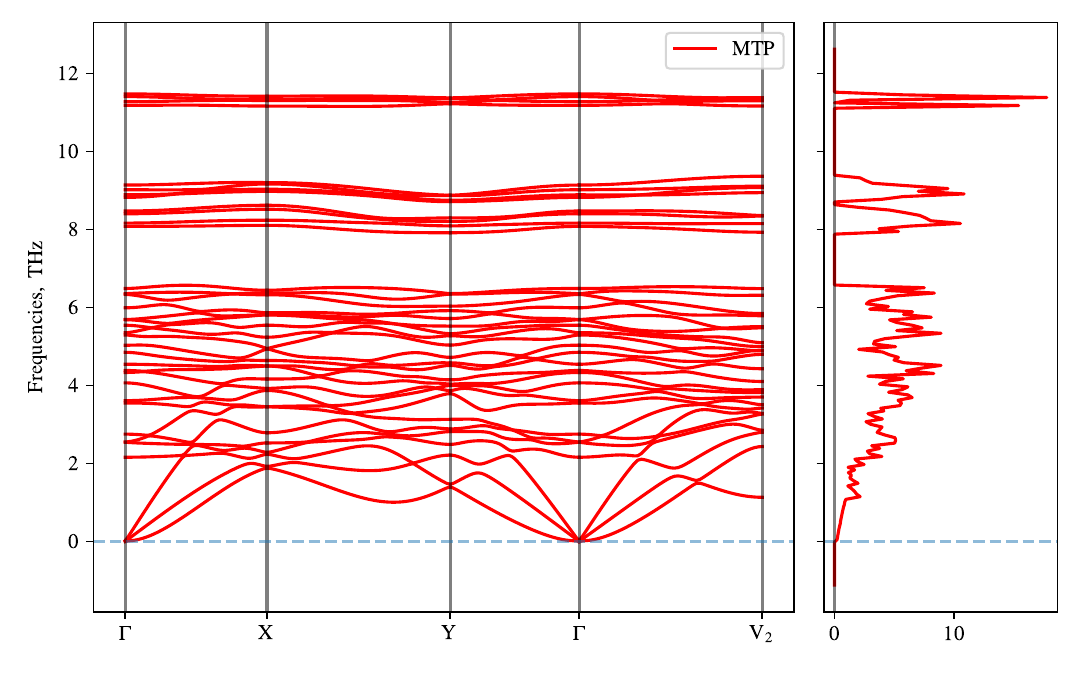}
        \caption{Mo$_2$S  }
    \end{subfigure}
      \begin{subfigure}[h!]{0.49\textwidth}
        \centering
        \includegraphics[width=\textwidth]{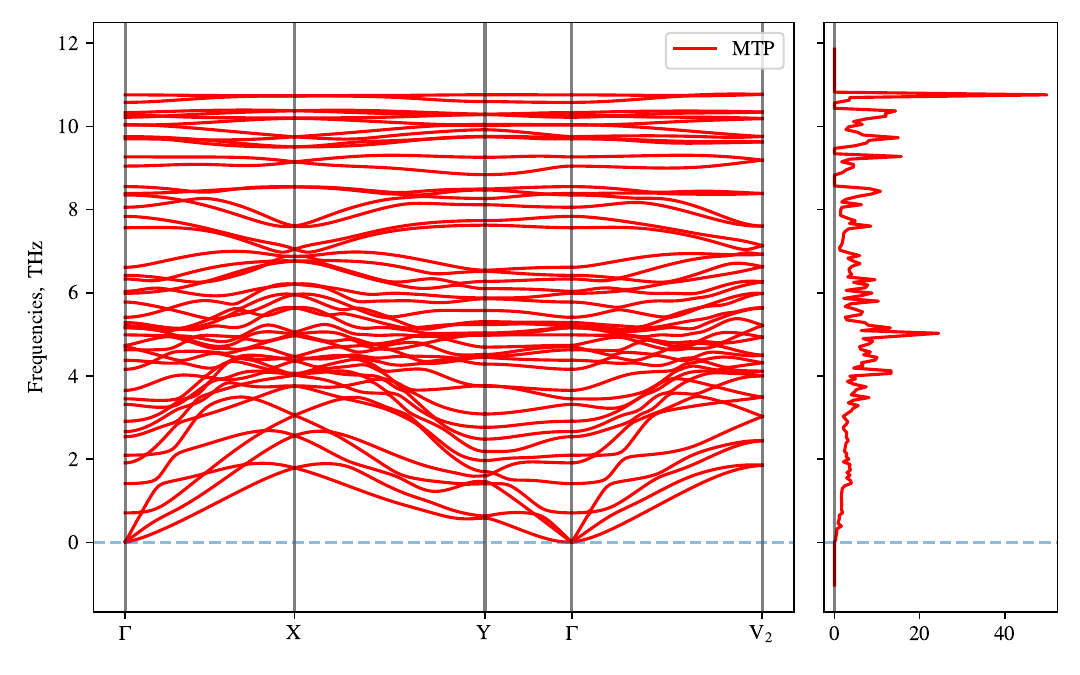}
        \caption{Mo$_5$S$_3$  }
    \end{subfigure}

    \begin{subfigure}[h!]{0.49\textwidth}
        \centering
        \includegraphics[width=\textwidth]{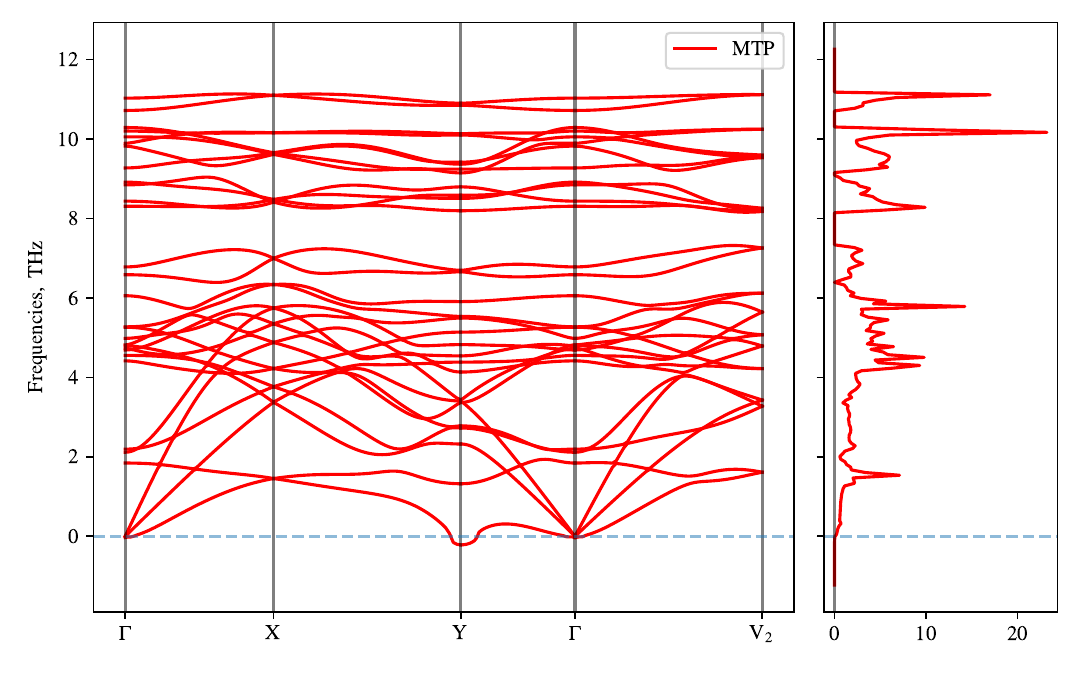}
        \caption{Mo$_3$S$_2$  }
    \end{subfigure}
     \begin{subfigure}[h!]{0.49\textwidth}
        \centering
        \includegraphics[width=\textwidth]{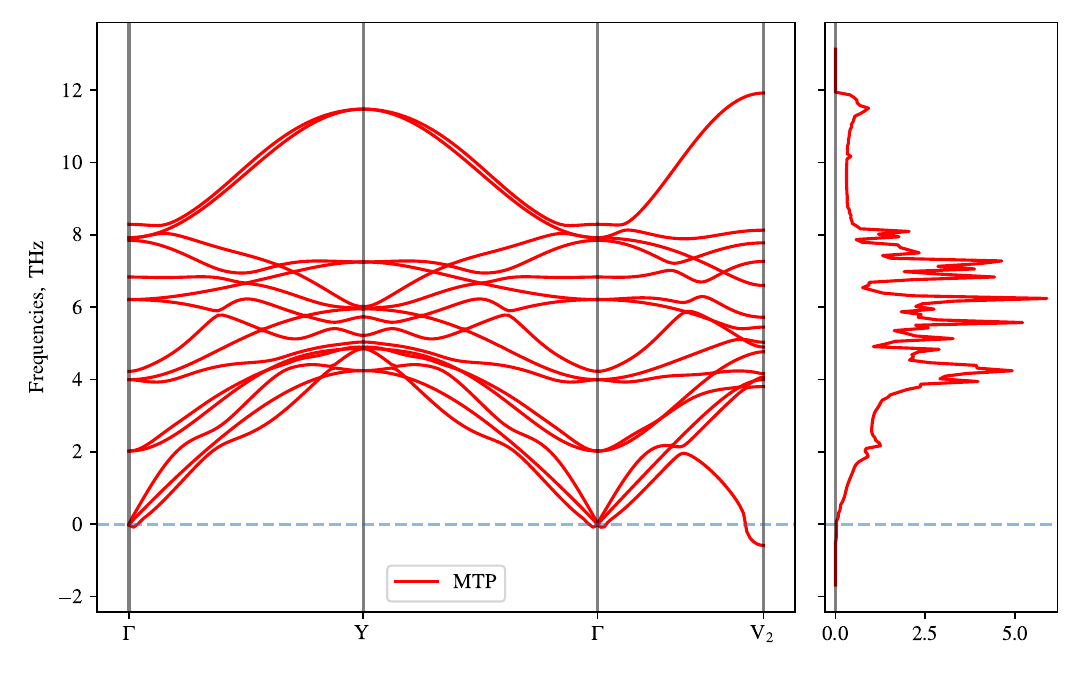}
        \caption{Mo$_4$S}
    \end{subfigure}
    \caption{Phonon band structures and phonon densities of states of the stable freestanding 2D crystals found during the evolutionary search, calculated with the MTP potential in the quasi-harmonic approximation using a 3x3x1 supercell. Almost all structures demonstrate no imaginary frequencies, which indicates their dynamical stability. In the case of Mo$_3$S$_2$, there is a Y-point instability. Nevertheless, the corresponding density of states is close to zero, so the overall contribution of this mode is small. }
        \label{fig:phonon_bs}
\end{figure*}

\section{Comparison of the DFT and MTP phonon bandstructures}\label{sec:phonon_bs_dft_mtp}

\begin{figure*}[h]
    \centering
    \begin{subfigure}[h!]{0.49\textwidth}
        \centering
        \includegraphics[width=\textwidth]{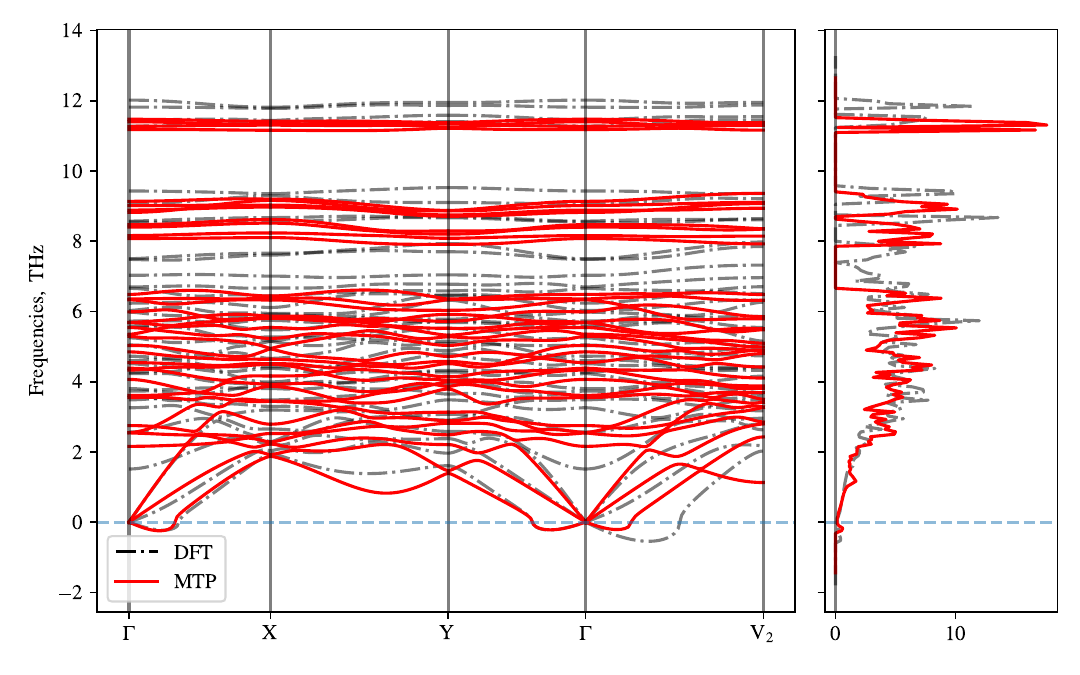}
        \caption{Mo$_2$S  }
    \end{subfigure}
      \begin{subfigure}[h!]{0.49\textwidth}
        \centering
        \includegraphics[width=\textwidth]{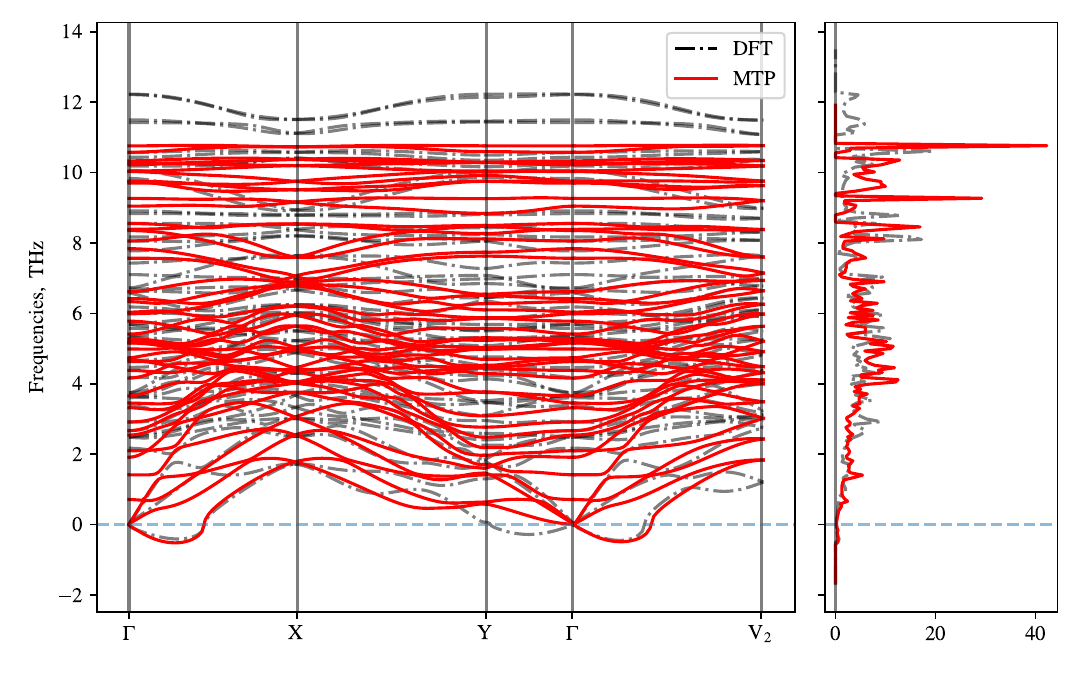}
        \caption{Mo$_5$S$_3$  }
    \end{subfigure}
    \begin{subfigure}[h!]{0.49\textwidth}
        \centering
        \includegraphics[width=\textwidth]{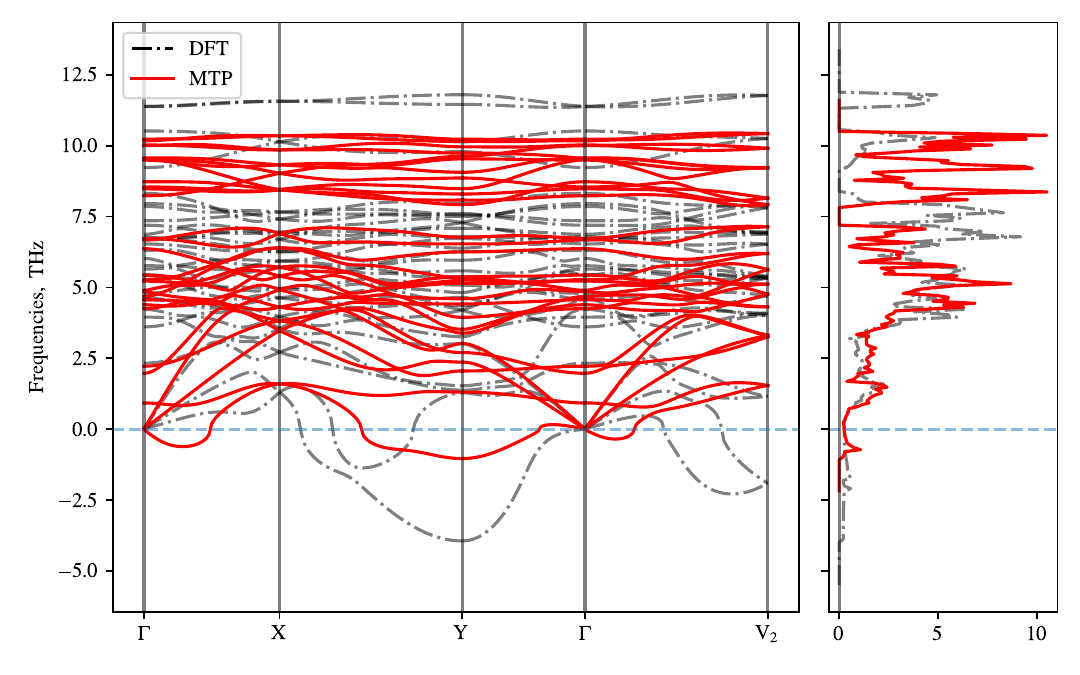}
        \caption{Mo$_3$S$_2$  }
    \end{subfigure}
     \begin{subfigure}[h!]{0.49\textwidth}
        \centering
        \includegraphics[width=\textwidth]{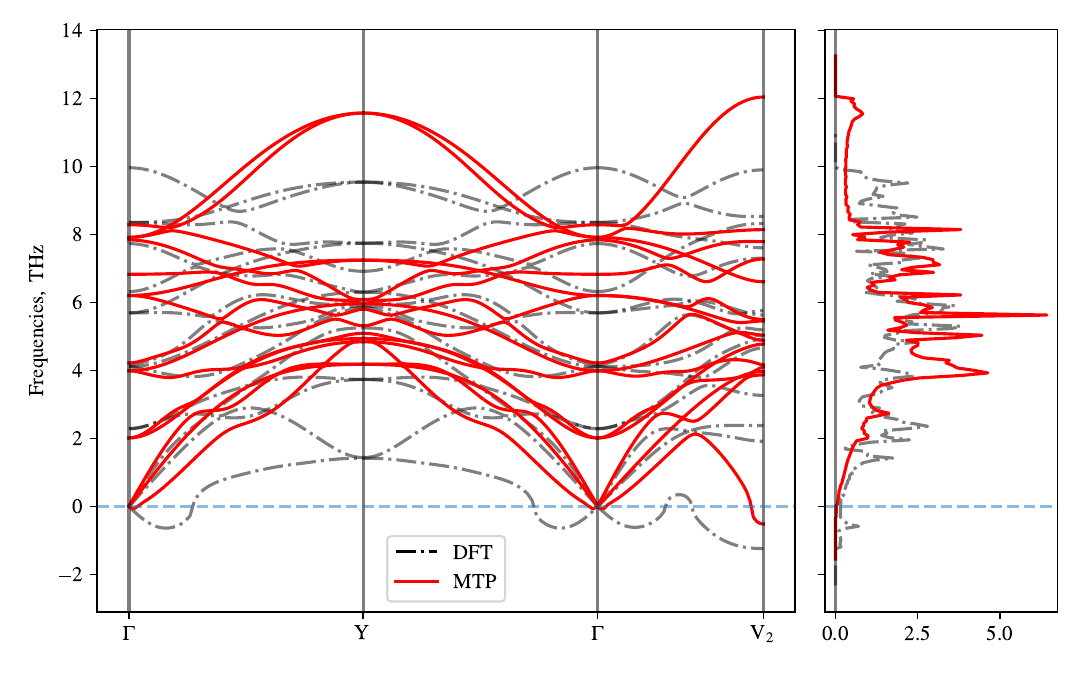}
        \caption{Mo$_4$S}
    \end{subfigure}
    \caption{Phonon band structures and phonon state densities of the stable free-standing 2D crystals calculated in 2x2x1 supercells with MTP (red solid lines) and DFT (dashed gray lines). Both methods show the presence of imaginary modes caused by the choice of a small supercell. The use of a larger supercell (i.e. 3x3x1) helps to obtain more accurate results without imaginary modes. }
        \label{fig:phonon_bs_mtp_dft}
\end{figure*}

\section{Visualization of the imaginary modes of the Mo$_3$S$_2$}\label{sec:Mo3S2_imag_vis}

\begin{figure*}[h]
    \centering
    \begin{subfigure}[h!]{0.49\textwidth}
        \centering
        \includegraphics[width=\textwidth]{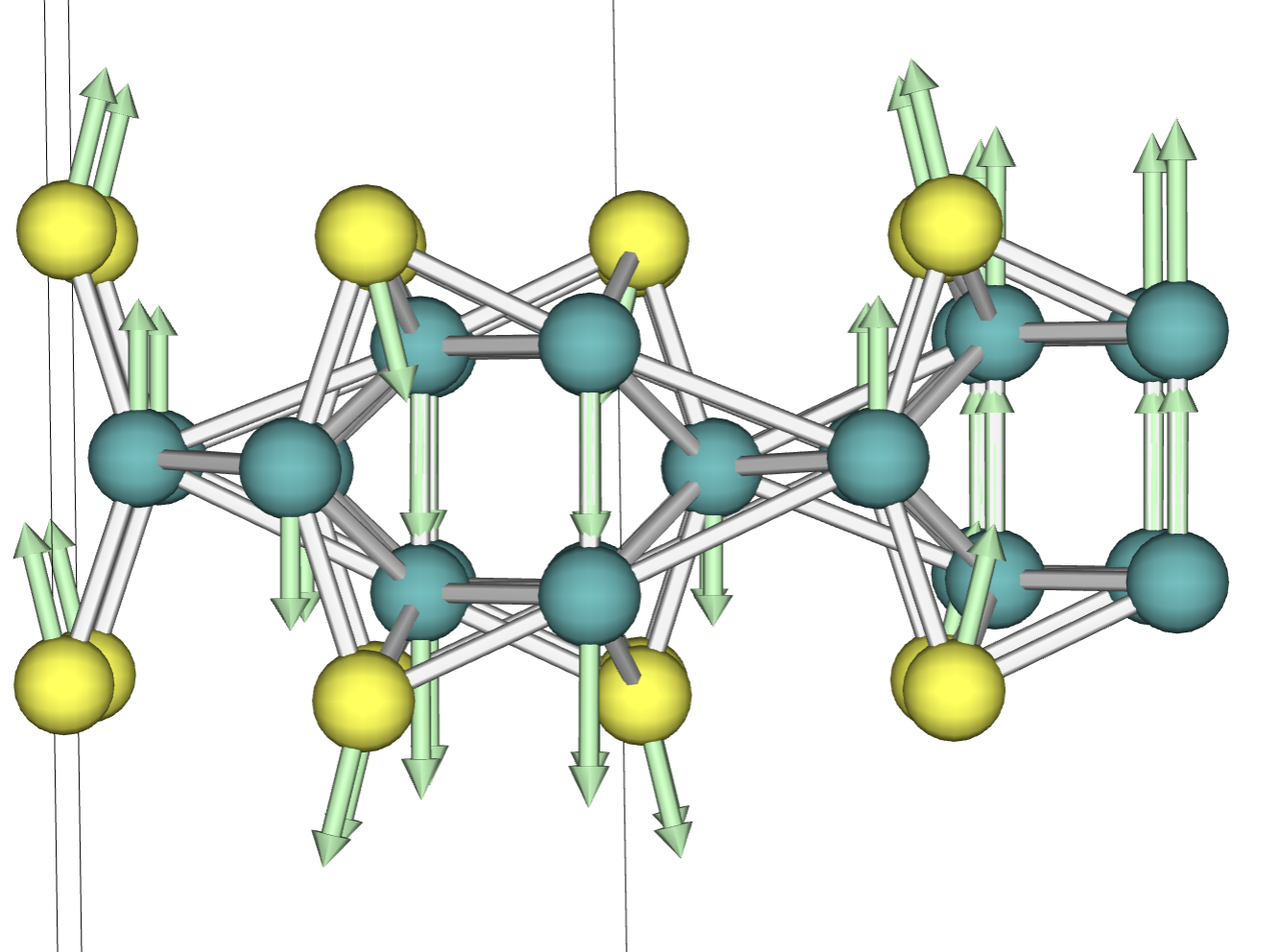}
        \caption{}
    \end{subfigure}
     \begin{subfigure}[h!]{0.49\textwidth}
        \centering
        \includegraphics[width=\textwidth]{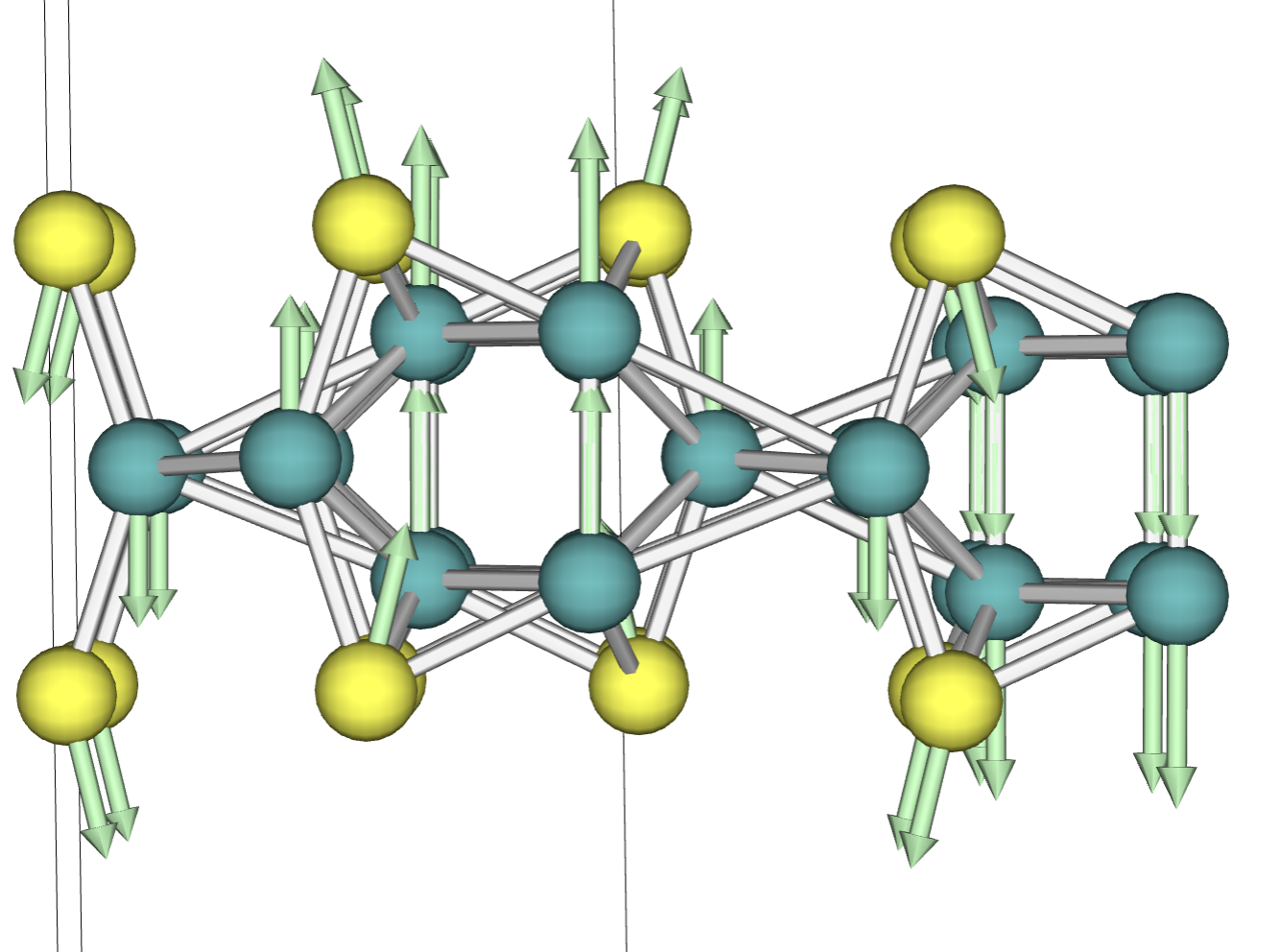}
        \caption{}
    \end{subfigure}
    
    \caption{Visualization of the vibration mode corresponding to the Y-point instability in the Mo$_3$S$_2$}
        \label{fig:Mo3S2_imag_vis}
\end{figure*}

\section{Anharmonicity effects in the phonon properties of the freestanding 2D MoS$_2$}\label{sec:phonon_anharm}

\begin{figure}[H]
    \centering
    \includegraphics[width=0.5\textwidth]{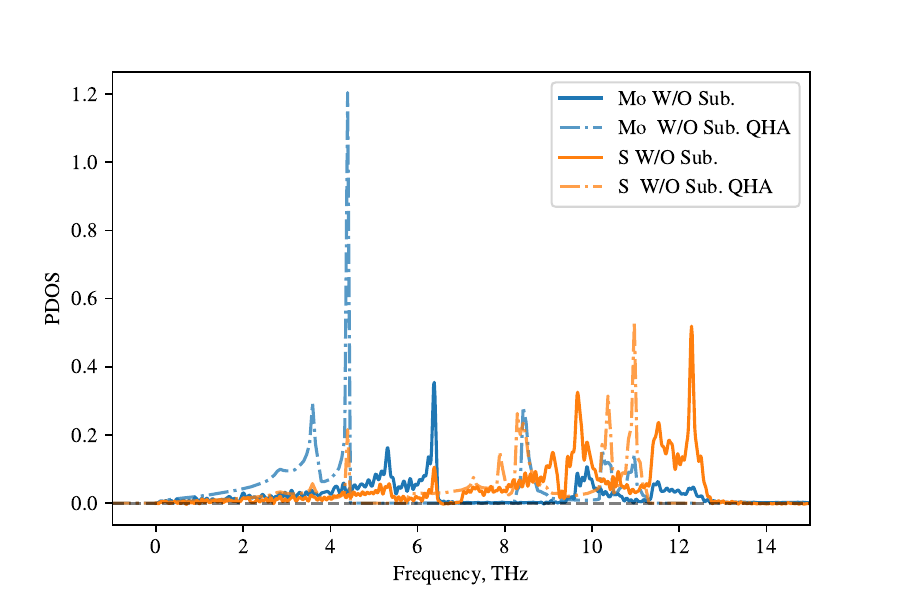}
    \caption{Phonon density of states of the free-standing 2D MoS$_2$ calculated with the MTP potential in the harmonic (dashed lines) and anharmonic (solid lines) setups. The partial contributions of molybdenum and sulfur are shown in blue and orange, respectively.}
    \label{fig:phonon_anharm}
\end{figure}

\section{Thermochemistry calculations details}

\subsection{Theoretical background}\label{sec:thermochemistry_theory}

All gases in our work have been considered in the ideal gas limit, assuming that the separation on translational, rotational and vibrational degrees of freedom is valid. Thus, the ideal gas enthalpy can be calculated as the sum of the internal energy at 0 K and the integral over the heat capacity at constant pressure:

\begin{equation}
    H(T)=E+E_{Z P E}+\int_0^T C_P d T
\end{equation}
where $E$ is a potential energy of the gas atom / molecule from the DFT calculation, $E_{ZPE}$ is a zero-point energy correction, and

\begin{equation}
    C_P=k_B+C_{V, \text { trans }}+C_{V, rot}+C_{V, vib}
\end{equation}
is separated into the translational, rotational and vibrational contributions. For a 3D gas, $C_{V, \text { trans }}$ is equal to $3/2 \ k_B$ , while $C_{V, rot}$  is 0 for monoatomic gas, $k_B$ for linear molecules and $3/2 \ k_B$ for nonlinear molecules. The vibrational heat capacity in its integrated form can be represented as the sum over $3N-5$ (or $3N-6$) vibrational degrees of freedom for linear (or nonlinear) molecules:

\begin{equation}
    \int_0^T C_{V, v i b} d T=\sum_i \frac{\epsilon_i}{e^{\frac{\epsilon_i}{k_B T}}-1}
\end{equation}

 where $\epsilon_i = \hbar \omega_i$   are the vibrational energies with the frequencies $\omega_i$.

The entropy of the ideal gas of molecules also consists of corresponding translational, rotational and vibrational contributions

\begin{equation}
    S(T, P)=S_{\text {trans }}\left(T, P_0\right)+S_{\text {rot }}\left(T, P_0\right)+S_{\text {vib }}\left(T, P_0\right)-k_B \ln P / P_0
\end{equation}
where

\begin{equation}
    \begin{aligned}
& S_{\text {trans }}\left(T, P_0\right)=k_B\left(\frac{5}{2}+\ln \left[\frac{\left(\frac{2 \pi M k_B T}{h^2}\right)^{\frac{3}{2}} k_B T}{P_0}\right]\right) \\
& S_{\text {rot }}(T)=\left\{\begin{array}{c}
0, \text { for monoatomic gas } \\
k_B\left(1+\ln \frac{8 \pi^2 I k_B T}{\sigma h^2}\right), \quad \text { for linear molecules } \\
k_B\left(\frac{3}{2}+\ln \frac{\sqrt{\pi I_A I_B I_C}}{\sigma}\left(\frac{8 \pi^2 k_B T}{h^2}\right)\right), \quad \text { for nonlinear molecules }
\end{array}\right. \\
& S_{v i b}(T)=k_B \sum_i\left(\frac{\epsilon_i}{k_B T\left(e^{\frac{\epsilon_i}{k_B T}}-1\right)}-\ln \left(1-e^{-\frac{\epsilon_i}{k_B T}}\right)\right) \\
&
\end{aligned}
\end{equation}

$I_A, I_B, I_C$ are the principal components of the inertia tensor of the nonlinear molecule, $I$ is the degenerate moment of inertia for a linear molecule and $\sigma$ is the symmetry number of the molecule determined from its point group symmetry.

\begin{figure*}[h]
    \centering
    \begin{subfigure}[h!]{0.33\textwidth}
        \centering
        \includegraphics[width=\textwidth]{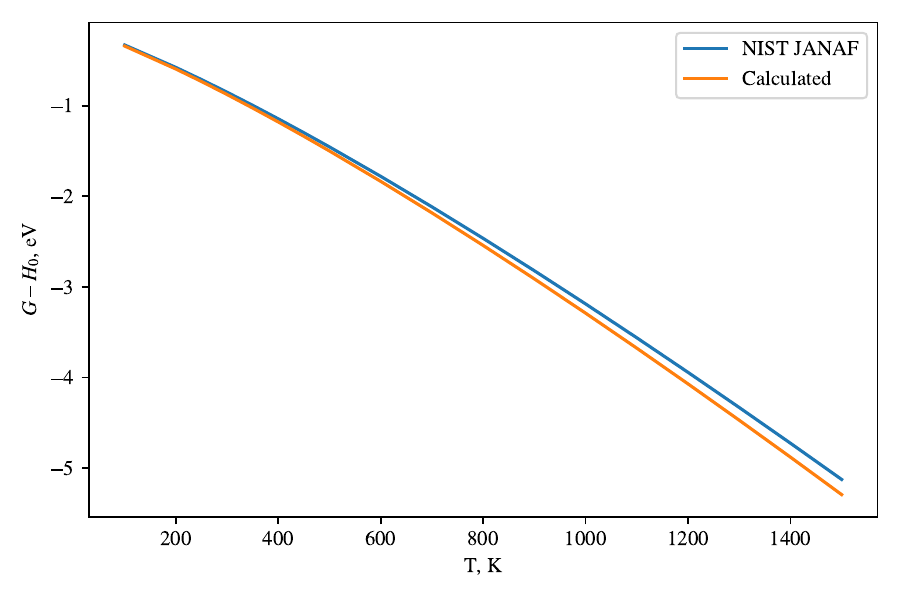}
        \caption{MoO$_3$}
    \end{subfigure}%
      \begin{subfigure}[h!]{0.33\textwidth}
        \centering
        \includegraphics[width=\textwidth]{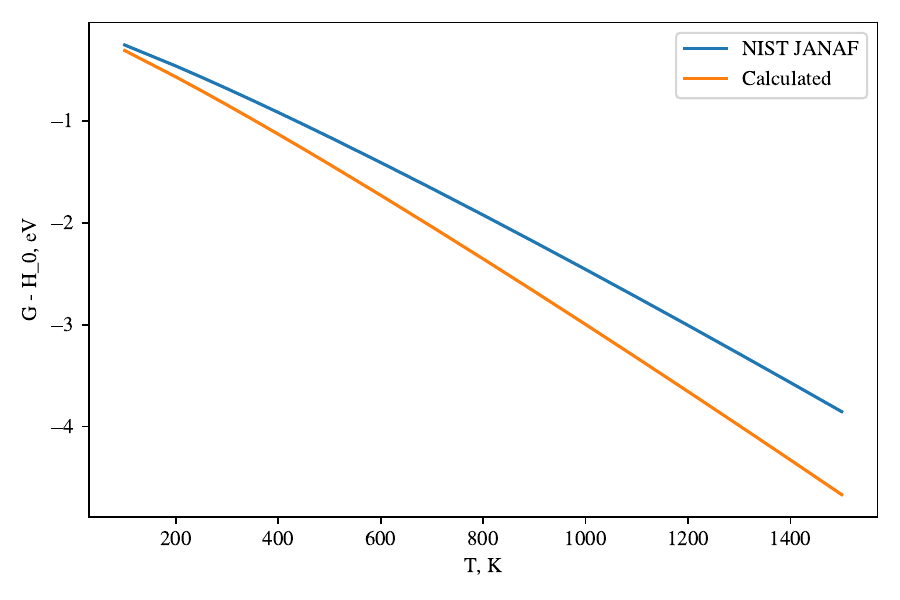}
        \caption{S$_2$}
    \end{subfigure}%
    \begin{subfigure}[h!]{0.33\textwidth}
        \centering
        \includegraphics[width=\textwidth]{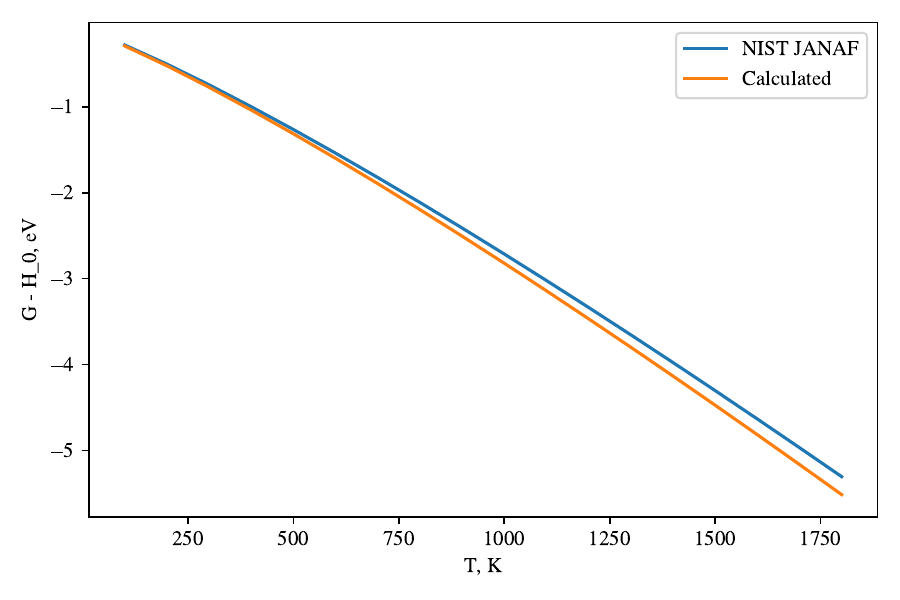}
        \caption{SO$_2$}
    \end{subfigure}
    \caption{Temperature dependence of the Gibbs free energy of MoO$_3$ (a), S$_2$ (b), and SO$_2$ (c) per molecule. Experimental results from NIST JANAF thermochemical tables. Theoretical calculations were performed using VASP and Atomic Simulation Environment. The reference energy values for the experimental data are made consistent with the DFT results.}
        \label{fig:thermo}
\end{figure*}

The dependence of the chemical potential on temperature and pressure can be obtained as the Gibbs free energy per molecule in the ideal gas approximation:

\begin{equation}\label{eq:chemical_potential}
    \mu(T, P)=G(T, P)=H(T)-T S(T, P)
\end{equation}
For each gas considered in our work, we calculated the values of $G(T, P)$ at standard pressure of $P = 1$ atm. and different temperatures according to the Eq. \eqref{eq:chemical_potential} and compared them with the experimental data from the NIST JANAF databases (Fig. \ref{fig:thermo}). One can see that in the given temperature range the MoO$_3$ and SO$_2$ precursors can be correctly described with the presented approach. The discrepancy between the theoretical and experimental results for $S_2$ gas is most likely caused by a rich chemistry of sulfur, which exhibits different molecular configurations in a gas phase (from S$_2$ to S$_8$ molecules), whose relative concentrations change with increase in temperature \cite{Rau1973ThermodynamicsVapour}. 

\subsection{Calculation of the chemical potential of sulfur}\label{sec:mu_S_theory}

\begin{figure}[H]
        \centering
        \includegraphics[width=0.5\textwidth]{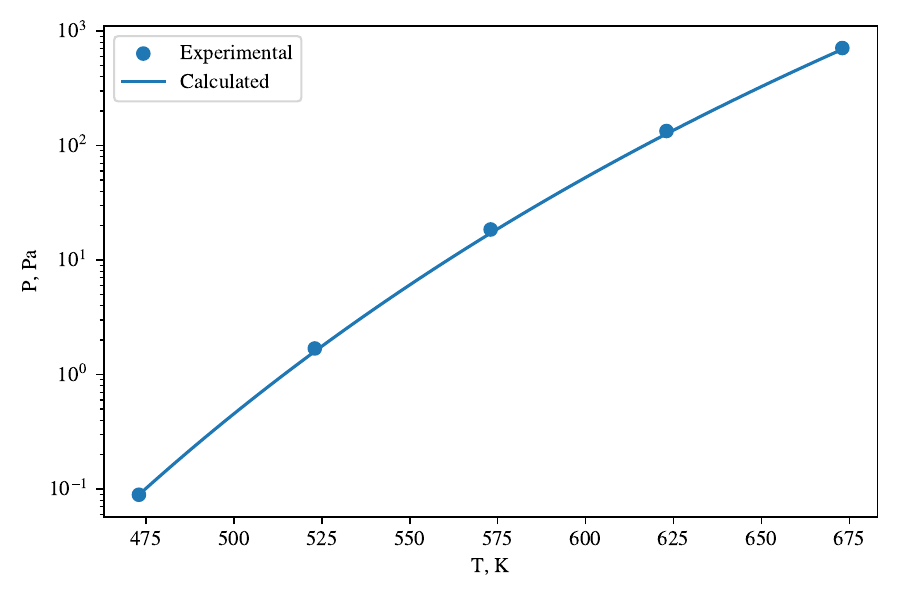}
        \caption{Saturated vapor pressure of sulfur (S$_2$) at different temperatures. Experimental data from Ref. \cite{Rau1973ThermodynamicsVapour}. Theoretical calculations are performed using the Clapeyron-Clausius equation with the fitted value of  the vaporization enthalpy $\Delta H_{vap} = 118.53 $ kJ/mol.}
        \label{fig:S2_vapor}
    \end{figure}

In the experimental setup, sulfur is evaporated from the container and then transferred to the reaction chamber with a carrier gas. Therefore, in the thermodynamic limit, we can estimate its pressure as the saturated vapor pressure at the temperature of the sulfur container. This value can be derived from the Clapeyron-Clausius equation, if a saturated vapor pressure at some temperature $T_0$ and pressure $P_0$ is known:

\begin{equation}\label{eq:Clapeyron-Clausius}
    \ln(P / P_0) = -\frac{\Delta H_{vap}}{R} (\frac{1}{T} - \frac{1}{T_0})
\end{equation}
where $\Delta H_{vap}$ is a vaporization enthalpy and $R$ is a universal gas constant. Uusing the reference data for the saturated vapor pressure of sulfur in its $S_2$ form at various temperatures from Ref. \cite{Rau1973ThermodynamicsVapour}, we fitted the value of $\Delta H_{vap} = 118.53$ kJ/mol. In Figure \ref{fig:S2_vapor}, we show a comparison of the theoretical values calculated using Eq. \eqref{eq:Clapeyron-Clausius} with the experimental results. 

Since all of the sulfur will eventually be transferred to a reaction chamber heated to $T=T_F$ - the furnace temperature, the chemical potential of sulfur can be calculated using the Eq. \eqref{eq:chemical_potential} by substituting the calculated value of the saturated vapor pressure and the furnace temperature:

\begin{equation}
    \mu_{S_2}(P, T) = \mu_{S_2}(P_{vap}(T_{SB}), T_F)
\end{equation}

\subsection{Calculation of the chemical potential of molybdenum}\label{sec:mu_Mo_theory}

The calculation of the chemical potential of molybdenum turns out to be a much more complicated task, since it is not present in a pure state in the experimental setup, and therefore there is no direct control over the parameters of its pressure and temperature. The actual chemistry of the sulfur-molybdenum oxide system is extremely rich and can exhibit many different intermediate steps (i.e., formation of intermediate MoS$_x$O$_y$ complexes) depending on the composition of the sulfur vapor and the external conditions \cite{Zhu2017CaptureMoS2}, thus making kinetics play a dominant role in the synthesis results. To obtain a quantitative thermodynamic approximation of the process, we assume that during the synthesis, MoO$_3$ first undergoes sublimation from the vessel, followed by a reaction with sulfur vapor, leading to the formation of molybdenum and sulfur dioxide:

\begin{equation}\label{eq:reaction}
     \mathrm{S_2(g) + MoO_3 (g) \rightarrow Mo (s) + SO_2(g)}.
\end{equation}

In the thermodynamic limit, we can therefore derive the $\mu_{Mo}$ as 

\begin{equation}\label{eq:mu_Mo}
    \mu_{Mo} = \frac{1}{4} [3 \mu_{S_2} + 4 \mu_{MoO_3} - 6 \mu_{SO_2}]
\end{equation}
where we balanced the equation to preserve the chemical composition.

Following the same reasoning as for the calculation of the chemical potential of sulfur, we estimated the partial pressure of MoO$_3$ vapor as the saturated vapor pressure using Eq. \eqref{eq:Clapeyron-Clausius}, replacing the enthalpy of vaporization by the enthalpy of sublimation. To find the value of the sublimation enthalpy, we used the experimental data from Ref. \cite{Viswanathan2014MolybdenumImplications}, and obtained $\Delta H_{sub} = 417.77$ kJ/mol

\begin{figure}[H]
        \centering
        \includegraphics[width=0.5\textwidth]{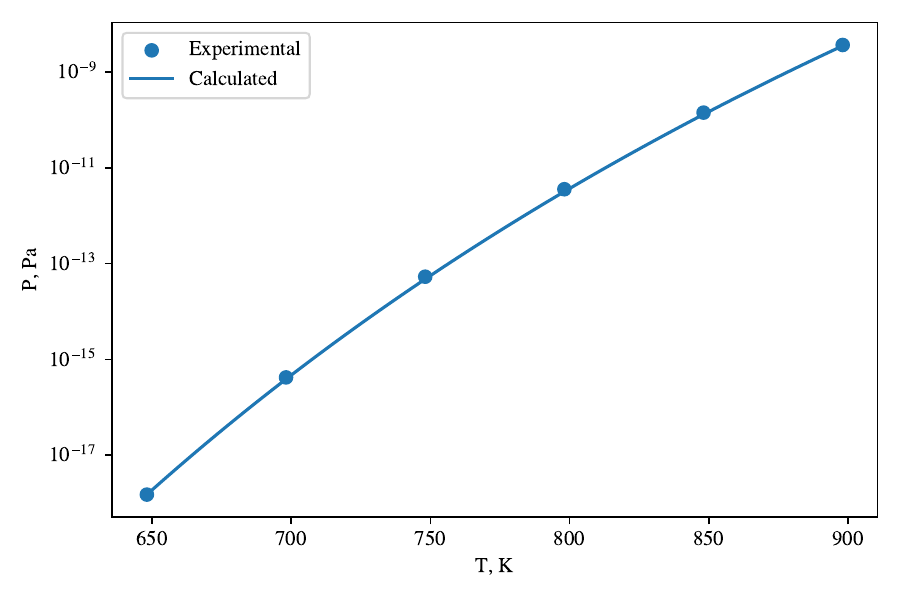}
        \caption{Saturated vapor pressure of MoO$_3$ at different temperatures. Experimental data from Ref. \cite{Viswanathan2014MolybdenumImplications}. Theoretical calculations are performed using the Clapeyron-Clausius equitation with the fitted value of  sublimation enthalpy $\Delta H_{sub} = 417.77$ kJ/mol }
        \label{fig:MoO3_vapor}
    \end{figure}

Once determined, the value of the saturated vapor pressure of  MoO$_3$ at $P = P_{vap}(T_F)$ and $T=T_F$ can be used to calculate the chemical potential of  MoO$_3$:

\begin{equation}
    \mu_{MoO_3}(P, T) = \mu_{MoO_3}(P_{vap}(T_{F}), T_F)
\end{equation}

Therefore, the final missing part is the chemical potential of $SO_2$ gas. To get its partial pressure, we used a formula for the equilibrium constant expressed in terms of partial pressures $K_P$:

\begin{equation}\label{eq:so2_pressure}
   K_P = \frac{(P_{SO_2})^6}{(P_{S_2})^3(P_{MoO_3})^4} = K_{eq} \cdot P^{6-3-4} = K_{eq} \cdot P^{-1}
\end{equation}
where $P_{SO_2}$, $P_{S_2}$ and $P_{MoO_3}$ are partial pressures of the reactants and products in gas phase, $K_{eq}$ is the equilibrium constant and $P$ is the total pressure of the reactants and products. The equilibrium constant can be derived from the Gibbs energy of reaction described in Eq. \eqref{eq:reaction}

\begin{equation}
    K_{eq} = e^{-\Delta G_r / k_B T}
\end{equation}
where $\Delta G_r$ at can be calculated using the Gibbs energies of formation of the components either from the thermochemical tables, or from the ideal gas formulas above. For simplicity, we calculated the value of $\Delta G_r$ at $P = 1$ atm., and $T=T_F$, and considered it constant for the rest of the calculations. The resulting value is $\Delta G_r = 75.645$ kJ/mol with the $K_{eq} = 1.1 \cdot 10^{-4}$.  

Finally, since the partial pressure of sulfur vapor at $T = T_F$ mostly dominates in the total pressure of reactants and products $P$ ($P_{vap, S_2}(T_F) \sim 19.6 \cdot 10^{4}$ Pa), we estimated P to be 0.2 atm. Therefore, the final value of $P_{SO_2}$ from Eq. \eqref{eq:so2_pressure} is $6.3 \cdot 10^{-4} $ Pa. 

Thus, by substituting of the values of partial pressures of SO$_2$, S$_2$ and MoO$_3$ at temperature $T=T_F$ to the ideal gas formulas, we can first derive the values of the chemical potentials of these components, and finally obtain the chemical potential of molybdenum from Eq. \eqref{eq:mu_Mo}. 

We note, that even though in this particular example the contribution of SO$_2$ is almost negligible, the overall theoretical result is still important for further studies, that may involve different experimental setups and conditions. 

\subsection{Dependence of the chemical potentials of molybdenum and sulfur on the experimental conditions}

\begin{figure*}[h]
    \centering
    \begin{subfigure}[h!]{0.49\textwidth}
        \centering
        \includegraphics[width=\textwidth]{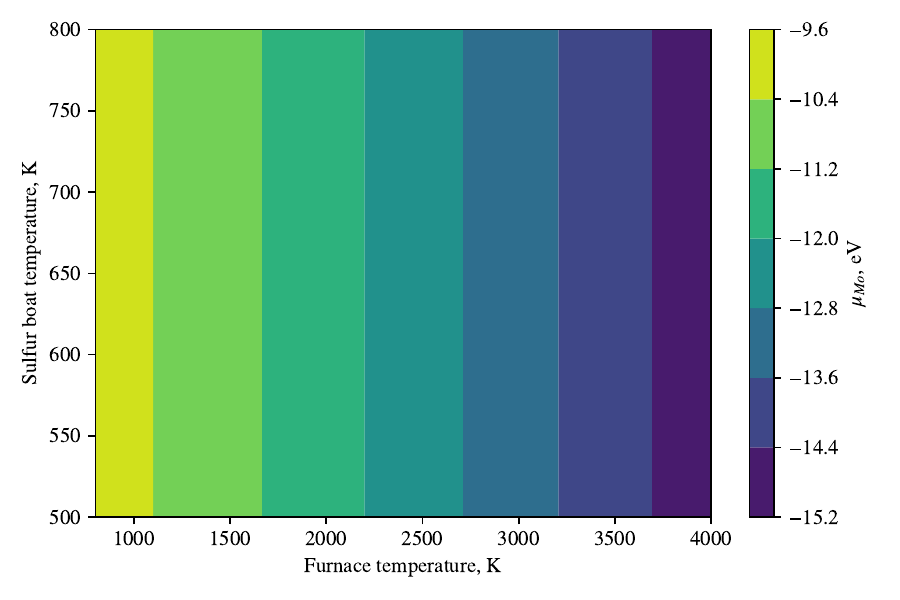}
    \end{subfigure}%
      \begin{subfigure}[h!]{0.49\textwidth}
        \centering
        \includegraphics[width=\textwidth]{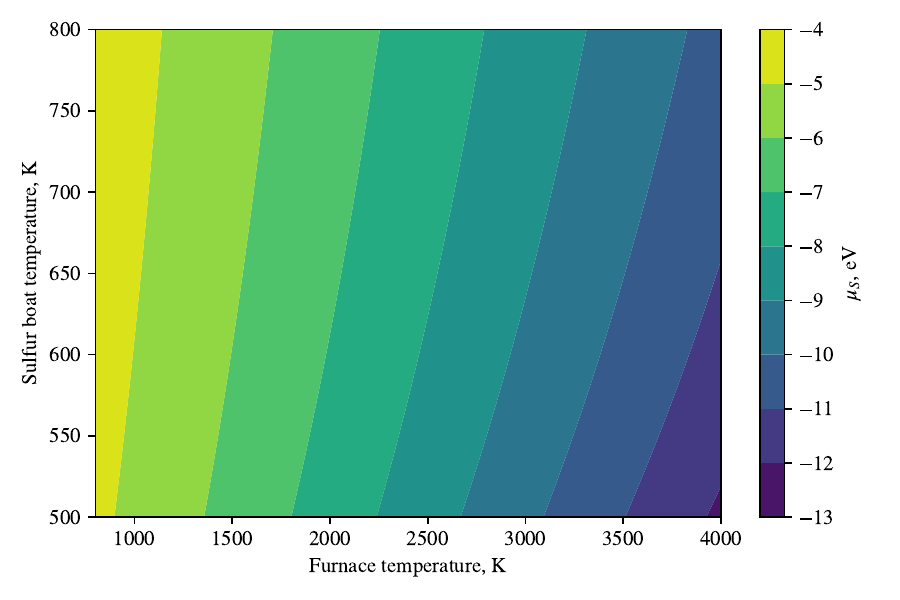}
    \end{subfigure}
    \caption{Values of chemical potentials of molybdenum (a) and sulfur (b) as a function of experimental parameters of CVD growth: furnace temperature and sulfur boat temperature.}
        \label{fig:chemical_potentials}
\end{figure*}

For demonstration purposes, in Figure \ref{fig:chemical_potentials}, we provide the dependence of the calculated chemical potentials $\mu_{Mo}$ and $\mu_{S}$ on the experimental conditions.

\subsection{Dependence of the stability map on the pressure of the components }

However in the text above we assume that the pressure of sulfur and MoO$_3$ in the setup is equal to their saturated vapor pressure, this is almost never true in reality. The presence of a constant flow of carrier gas through the reaction chamber increases the velocity of both sulfur and MoO$_3$ vapors and thus decreases their pressure according to Bernoulli's principle. The actual dependence of the reactant pressure on the carrier gas velocity depends strongly on the geometry of each specific experimental setup. Nevertheless, we can assume that different flow rates effectively scale the initial saturated pressure by a constant $\alpha < 1$, resulting in a following updated value of the pressure for each precursor:

\begin{equation}
    P = \alpha(\text{Ar flow rate}) \cdot P_{vap}.
\end{equation}

\begin{figure*}[h]
    \centering
    \begin{subfigure}[h!]{0.33\textwidth}
        \centering
        \includegraphics[width=\textwidth]{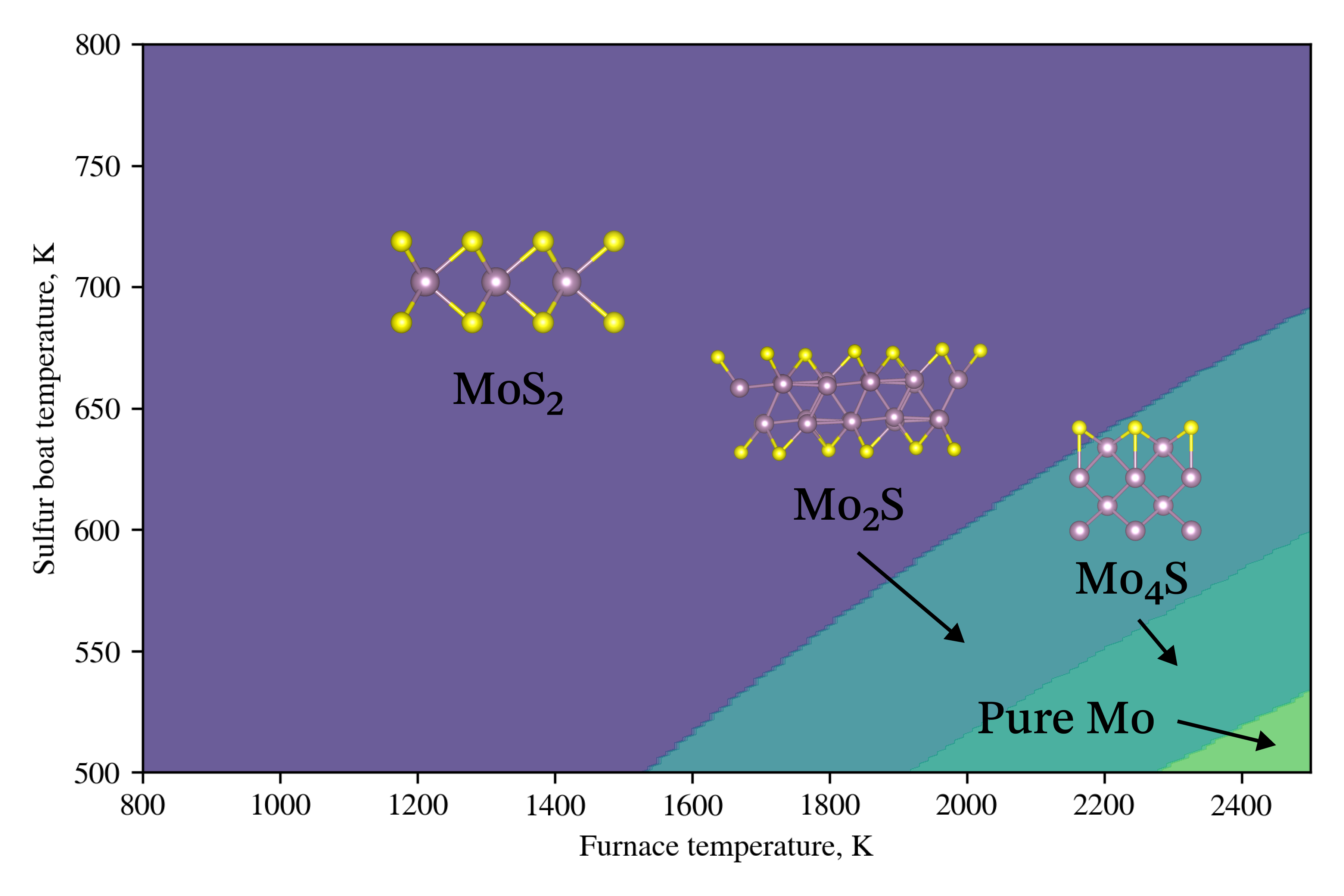}
        \caption{$\alpha=0.01$}
    \end{subfigure}%
      \begin{subfigure}[h!]{0.33\textwidth}
        \centering
        \includegraphics[width=\textwidth]{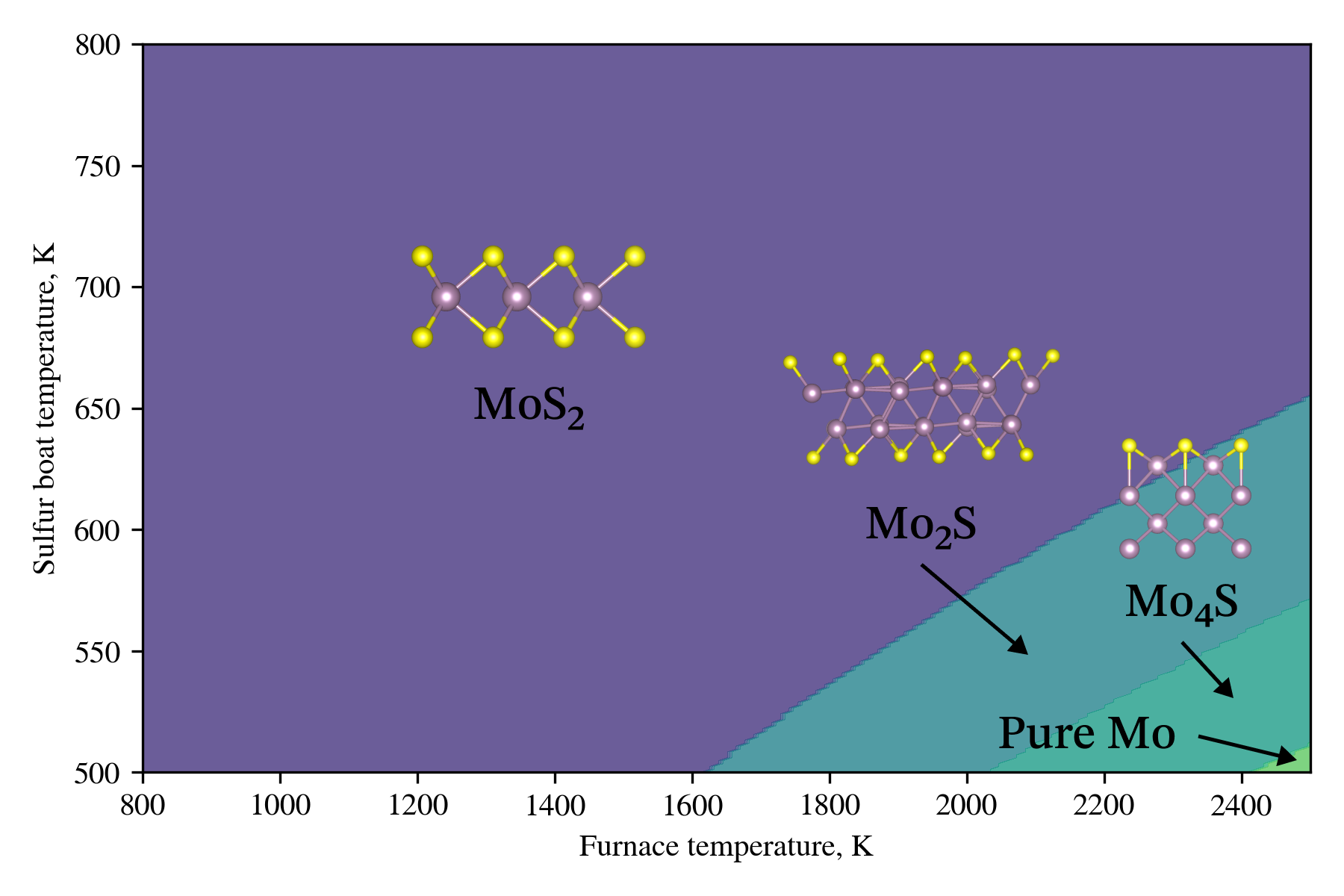}
        \caption{$\alpha=0.1$}
    \end{subfigure}%
    \begin{subfigure}[h!]{0.33\textwidth}
        \centering
        \includegraphics[width=\textwidth]{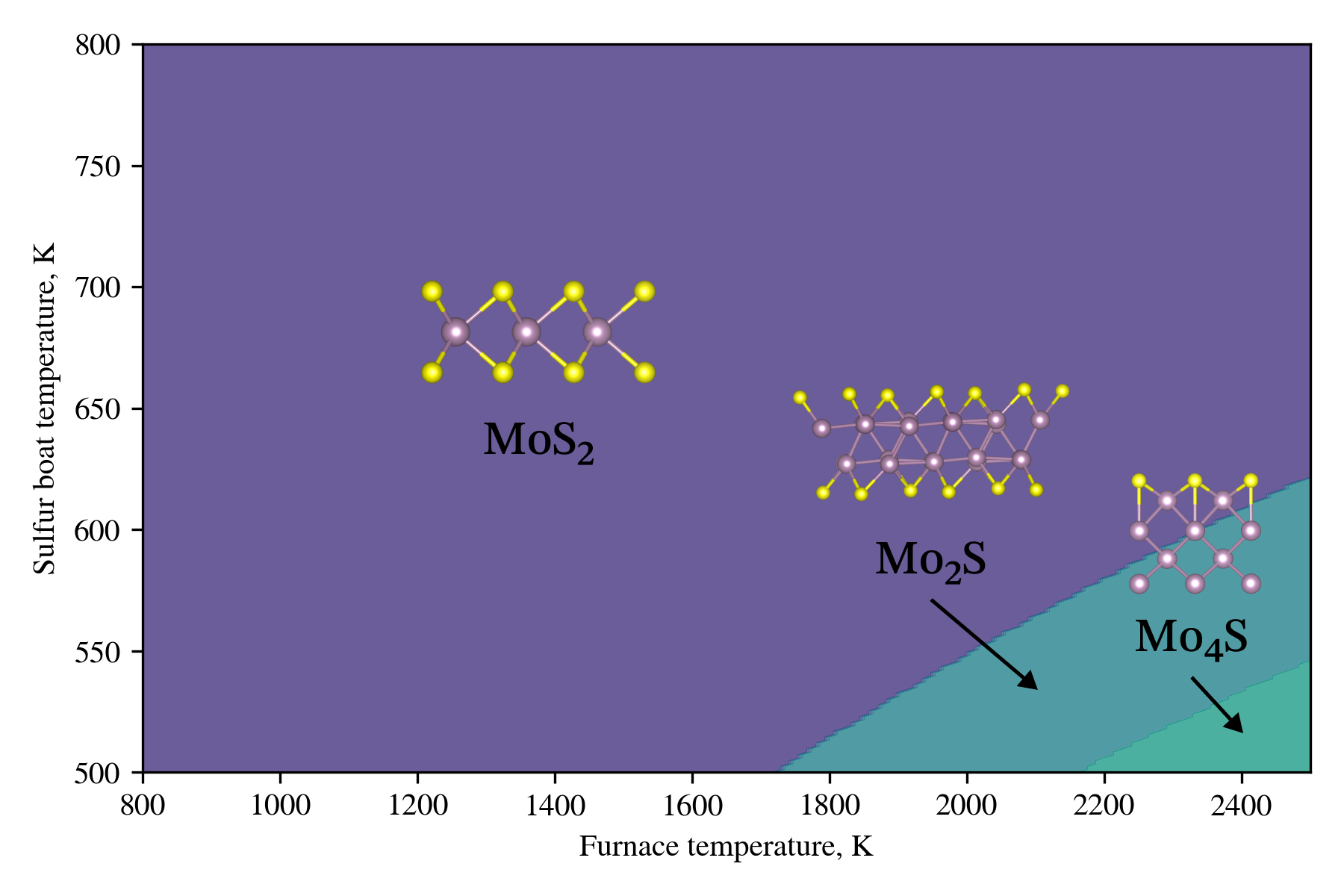}
        \caption{$\alpha=1.0$}
    \end{subfigure}
    \caption{Stability map of the 2D Mo-S system depending on the value of the pressure scaling coefficient $\alpha$. This coefficient effectively represents different flow rates of the carrier gas, which increases the velocity of the MoO$_2$ and S$_2$ vapors and decreases their pressure according to Bernoulli's principle. The value of $\alpha = 1.0$ means that the pressure of both vapors are equal the their saturated vapor pressure.}
        \label{fig:stability_maps_alpha}
\end{figure*}

Figure \ref{fig:stability_maps_alpha} shows how the stability map of the 2D Mo-S system changes with the change of the coefficient $\alpha$. Decreased pressure of the precursors leads to decrease in their chemical potential values and thus effectively shifts the stability regions to lower temperature range. This effect allows to achieve successful synthesis at more moderate temperatures, thus simplifying the synthesis procedure. 

\bibliography{si}% common bib file